\documentclass{nature}
\usepackage{amsmath}
\usepackage[T1]{fontenc}
\usepackage[latin1]{inputenc}
\usepackage{graphicx}
\usepackage{amssymb}
\usepackage{dcolumn}
\usepackage{color}
\usepackage{bm}
\usepackage[normalem]{ulem}
\usepackage{url}
\usepackage{chemformula}
\usepackage{pdfpages}
\usepackage{pgffor}

\begin{document}

\title{Determination of topological edge quantum numbers of fractional quantum Hall phases}

\author{Saurabh Kumar Srivastav$^{1}$, Ravi Kumar$^{1}$, Christian Sp{\r a}nsl{\"a}tt$^{2}$, K. Watanabe$^{3}$, T. Taniguchi$^{3}$, Alexander D. Mirlin$^{4,5,6,7}$, Yuval Gefen$^{4,8}$, and Anindya Das$^{1}$\footnote{anindya@iisc.ac.in}}

\maketitle

\begin{affiliations}

\item Department of Physics, Indian Institute of Science, Bangalore, 560012, India.
\item Department of Microtechnology and Nanoscience (MC2),Chalmers University of Technology, S-412 96 G\"oteborg, Sweden.
\item National Institute of Material Science, 1-1 Namiki, Tsukuba 305-0044, Japan.
\item Institute for Quantum Materials and Technologies, Karlsruhe Institute of Technology, 76021 Karlsruhe, Germany.
\item Institut f{\"u}r Theorie der Kondensierten Materie, Karlsruhe Institute of Technology, 76128 Karlsruhe, Germany.
\item Petersburg Nuclear Physics Institute, 188300 St. Petersburg, Russia.
\item L. D. Landau Institute for Theoretical Physics RAS, 119334 Moscow, Russia.
\item Department of Condensed Matter Physics, Weizmann Institute of Science, Rehovot 76100, Israel.

\end{affiliations}

\noindent
\textbf{To determine the topological quantum numbers of fractional quantum Hall (FQH) states hosting counter-propagating (CP) downstream ($N_d$) and upstream ($N_u$) edge modes, it is pivotal to study quantized transport both in the presence and absence of edge mode equilibration. While reaching the non-equilibrated regime is challenging for charge transport, we target here the thermal Hall conductance $G_{Q}$, which is purely governed by edge quantum numbers $N_d$ and $N_u$. Our experimental setup is realized with a hBN encapsulated graphite gated monolayer graphene device. For temperatures up to $35mK$, our measured $G_{Q}$ at $\nu = $ 2/3 and 3/5 (with CP modes) match the quantized values of non-equilibrated regime $(N_d + N_u)\kappa_{0}T$, where $\kappa_{0}T$ is a quanta of $G_{Q}$. With increasing temperature, $G_{Q}$ decreases and eventually takes the value of equilibrated regime $|N_d - N_u|\kappa_{0}T$. By contrast, at $\nu = $1/3 and 2/5 (without CP modes), $G_Q$ remains robustly quantized at $N_d\kappa_{0}T$ independent of the temperature. Thus, measuring the quantized values of $G_{Q}$ at two regimes, we determine the edge quantum numbers, which opens a new route for finding the topological order of exotic non-Abelian FQH states.
}

\noindent\textbf{Introduction.}
In the quantum Hall (QH) regime, transport occurs in one-dimensional gapless edge modes, which reflect the topology of  the bulk filling factor $\nu$. In integer QH (IQH) states and in a certain subclass of fractional QH (FQH) states, only downstream edge modes ($N_d$ of them) exist, whose chirality is dictated by the direction of the applied magnetic field\cite{beenakker1990edge,wen1990chiral}. At the same time, the edge structure of a majority of FQH states, including, in particular, the ``hole-like'' states $(1/2<\nu<1)$, is more complicated. In addition to the downstream edge modes, the presence of upstream modes ($N_u$) leads to complex transport behaviour\cite{beenakker1990edge,wen1990chiral,PhysRevLett.64.220,johnson1991composite,wen1992theory,Kane1994}. In this situation, the measured values of the electrical conductance ($G_{e}$) depends on the extent of the charge equilibration between the counter-propagating downstream and upstream modes. For example, the $\nu=2/3$ state  hosts two counter-propagating modes: a downstream mode, $\nu = 1$, and an upstream $\nu = 1/3$ mode\cite{PhysRevLett.64.220}. With full charge equilibration, the two-terminal conductance $G_{e}$ becomes\cite{kane1995impurity,protopopov2017,Nosiglia2018,spaanslatt2021contacts} $2e^2/3h$; on the other hand, in the absence of charge equilibration, $G_{e}$ is equal to\cite{protopopov2017,spaanslatt2021contacts} $4e^2/3h$. The observation of a crossover from $4e^2/3h$ to $2e^2/3h$ is essential to establish the proposed edge structure. This crossover has indeed been observed in carefully engineered double-quantum-well structure, allowing control of the equilibration\cite{Cohen2019}. At the same time, a similar demonstration is lacking in experiments on a conventional edge (the boundary of a $\nu = 2/3$ FQH state), where $G_{e}$ is always found to be $2e^2/3h$. The reason is that the small value of the charge equilibration length makes it difficult to access the non-equilibrated regime. A small deviation from $2e^2/3h$ indicating a beginning of the crossover towards $4e^2/3h$ was observed for the spin-unpolarized $\nu = 2/3$ FQH state\cite{lafont2019counter}.

\begin{figure*}
\centerline{\includegraphics[width=1\textwidth]{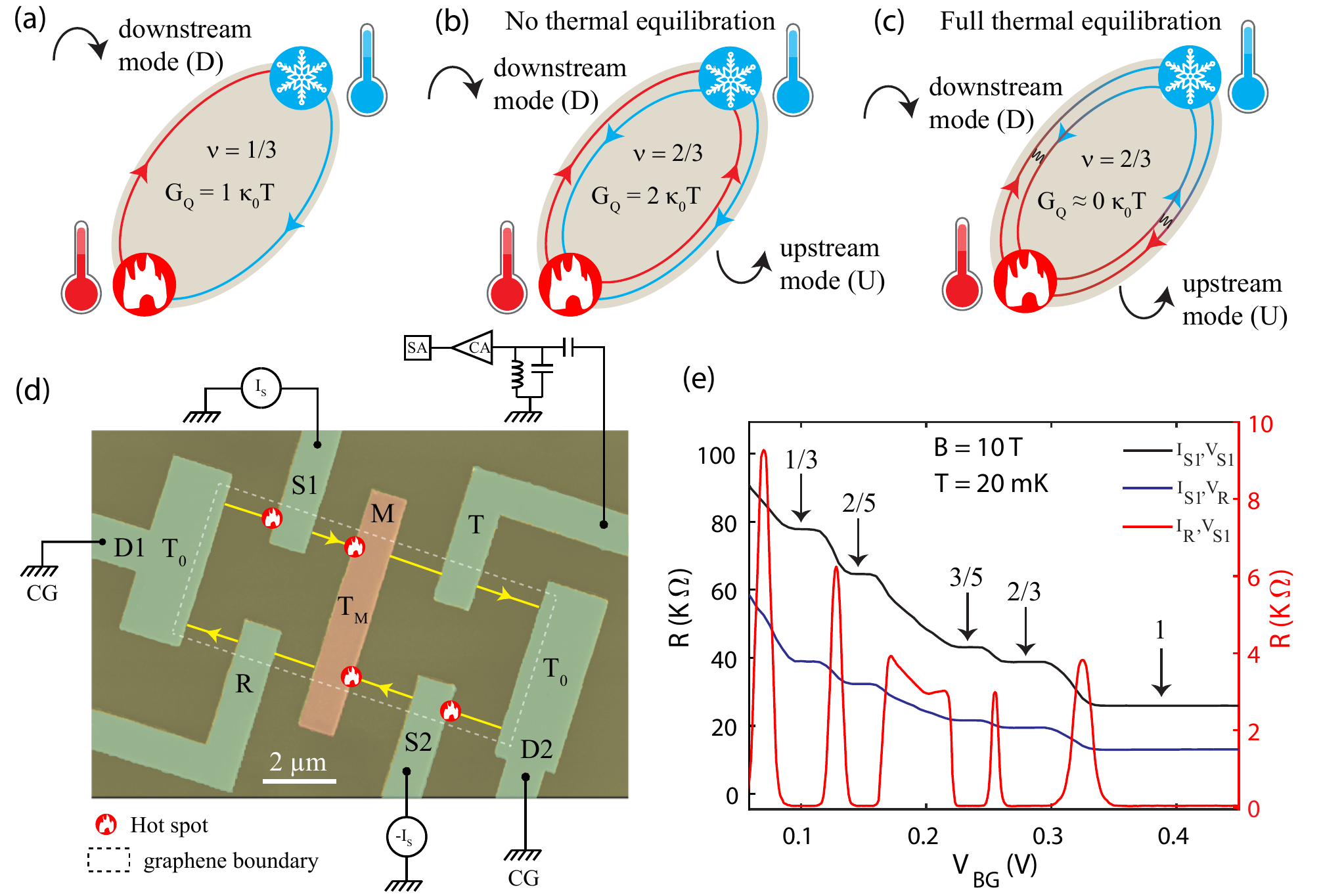}}
\caption{\textbf{Schematics of heat transport on QH edges, measurement setup, and QH response of device.} (\textbf{a}) Heat transport at the edge of $\nu = 1/3$ state along a single downstream mode. The chirality of the downstream mode is clockwise. (\textbf{b}) Heat transport at the edge of $\nu = 2/3$ state in non-equilibrated regime. Heat from the hot reservoir is carried away by both downstream and upstream modes. The chirality of upstream mode is anti-clockwise. (\textbf{c}) Heat transport at the edge of $\nu = 2/3$ state in equilibrated regime. The gradient of the color along the edges represents the qualitative temperature profile. In the long-length limit ($L \to \infty$), the heat carried away from the hot reservoir comes back to it via other edge modes, which leads to a vanishing thermal conductance. (\textbf{d}) False colored SEM micrograph of the device, shown with the measurement schematic. The graphene boundary is marked with a white dashed line. For illustrative purposes, the device is depicted with a $\nu$=1 edge structure. For thermal conductance measurements, currents $I_{S}$ and $-I_{S}$ are fed simultaneously at contacts $S1$ and $S2$. Due to the power dissipation near the central, floating contact, the electron temperature increases to $T_{M}$. The electrical and thermal conductances are measured respectively at low frequency (23 Hz) and high frequency ($\sim$ 740kHz) with an LCR resonant circuit. (\textbf{e}) QH response: The black line is the resistance $R_{S1}$ ($V_{S1}/I_{S1}$) measured at source contact `$S1$' as a function of $V_{BG}$ at B = 10T and temperature 20 mK. The blue line shows the measured resistance ($V_{R}/I_{S1}$) at the `$R$' contact. The red curve shows the resistance $V_{S1}/I_{R}$ measured at the contact `$S1$', while the current is injected at the contact `$R$' and encodes the longitudinal resistance. Robust fractional plateaus at $\frac{1}{3}\frac{e^2}{h}$, $\frac{2}{5}\frac{e^2}{h}$, $\frac{3}{5}\frac{e^2}{h}$, and $\frac{2}{3}\frac{e^2}{h}$ are clearly visible. The legend defines the current sources and voltage probes for each curve. The subscripts of $I$ and $V$ correspond to the current-fed contact and the voltage-probe contact, respectively.}
\label{Figure1}
\end{figure*}

Measurements of  thermal conductance have recently emerged as a powerful tool to detect the edge structure of FQH states\cite{kane1996thermal,kane1997quantized,banerjee2017observed,banerjee2018observation,Srivastaveaaw5798,PhysRevLett.126.216803}. Such measurements are highly useful for ``counting'' edge modes and can also detect charge neutral Majorana modes\cite{banerjee2018observation,kasahara2018majorana}. For IQH states and FQH states with only downstream modes, the quantized thermal conductance is given by $G_{Q}=N_{d}\kappa_{0}T$, where $\kappa_{0}= \pi^{2}k^{2}_\mathrm{B}/3h$, $k_\mathrm{B}$ is the Boltzmann constant, $h$ is the Planck constant, and $T$ is the temperature\cite{kane1997quantized}. A schematic illustration of the heat flow for such a state  ($\nu = 1/3$ in this example) is depicted in Fig.~1(a). On the other hand, for hole-like FQH states,  the presence of upstream modes renders the value of $G_{Q}$ strongly dependent on the extent of thermal equilibration between CP modes. This  leads to a crossover\cite{protopopov2017} of $G_{Q}$ from a non-equilibrated quantized value of $(N_{d}+N_{u})\kappa_{0}T$ to the asymptotic value of full equilibration  $|N_{d}-N_{u}|\kappa_{0}T$. Such a crossover behaviour of heat conductance is schematically shown in Fig.~1(b,c) for $\nu = 2/3$. While the fully-equilibrated and non-equilibrated limiting cases of $G_{Q}$ have been reported in disparate GaAs/AlGaAs based 2DEG devices\cite{banerjee2017observed,banerjee2018observation,melcer2022}, and in graphene only the non-equilibrated values have been  observed\cite{PhysRevLett.126.216803},  a crossover of $G_{Q}$ from the non-equilibrated to the  fully equilibrated limit in a single device has remained unattainable. This has remained  one of the long-standing challenges on the path  to reveal the detailed edge structure of FQH states. Achieving this goal would further help to settle the topological order of more complex non-Abelian even-denominator FQH states and may be useful for revealing possible reconstruction of QH edges.  

In the present work we report on  thermal conductance measurements (as a function of temperature $T$) of FQH states without CP modes ($\nu = 1/3$ and $2/5$) and with CP modes ($\nu = 2/3$ and $3/5$), realized in a $hBN$ encapsulated graphite gated high-mobility monolayer graphene device. Our key findings are the following: (1) At the base temperature ($\sim 20mK$),  $G_{Q}$ for $2/3$ and $3/5$ is found to be $2\kappa_{0}T$ and $3\kappa_{0}T$, respectively, which matches  the non-equilibrated limit $(N_{d}+N_{u})\kappa_{0}T$. These values remain constant up to $\sim 35mK$. (2) With further increase of temperature,  $G_{Q}$ for $3/5$ decreases, saturating at the  equilibrated limit  $|N_{d}-N_{u}|\kappa_{0}T = 1\kappa_{0}T$  for $T \gtrsim 50mK$. The crossover from the non-equilibrated to the equilibrated regime of $G_{Q}$ is observed for $2/3$ too. In this case, the heat transport in the equilibrated regime is of diffusive character, with the limiting value $|N_{d}-N_{u}|\kappa_{0}T \approx 0$ that is approached in a power-law way as a function of temperature. Our measurements show a drop of $G_{Q}$ that reaches a value $\sim 0.5\kappa_{0}T$ at $60mK$, continuing to decrease towards zero. (3) For $1/3$ and $2/5$ FQH states (no CP modes), $G_{Q}$ is found to be $1\kappa_{0}T$ and $2\kappa_{0}T$, respectively,  independent of the electron temperature.

\noindent\textbf{Device schematic and response:} 
To measure the thermal conductance, we have used a graphite-gated graphene device, where the graphene is encapsulated between two hBN layers. The details of the device fabrication is described in Methods. Similar to our previous work~\cite{Srivastaveaaw5798,PhysRevLett.126.216803}, our device consists of a small floating metallic reservoir, which is connected to graphene channel via one-dimensional edge contacts, as shown in Fig.~1(d). 
To measure the electrical conductance, we used the standard lock-in technique whereas the thermal conductance measurement was performed with noise thermometry~\cite{jezouin2013quantum,banerjee2017observed,banerjee2018observation,Srivastaveaaw5798,PhysRevLett.126.216803,kumar2022observation} (see SI). In Fig. 1(e), the black curve represents the measured resistance $R_{S1}$ ($V_{S1}/I_{S1}$) at the source contact ($S1$) as a function of the graphite gate voltage ($V_{BG}$). Well developed plateaus appear at $\nu =$ $\frac{1}{3}$, $\frac{2}{5}$, $\frac{3}{5}$ and $\frac{2}{3}$. The blue curve shows the measured resistance $R_{R} = V_{R}/I_{S1}$ along the reflected path (at contact $R$) from the floating contact. Measured resistances along the reflected path is exactly half of the resistance measured at the source contact, suggesting equal partitioning of injected current to both the transmitted and reflected side (see SI). The red curve in Fig. 1(e) shows the resistance $R_{S1} = V_{S1}/I_{R}$ measured at contact $S1$, while the current is injected from the contact $R$. This resistance in this configuration has the same properties as a longitudinal resistance: in the absence of bulk transport, the voltage $V_{S1}$ is determined by the equilibrium potential of the ground contact D1. The observation of the vanishing resistance plateaus further supports the formation of well developed FQH states. It should be noted that the measured resistance values suggest full charge equilibration in our device (see SI).

\begin{figure*}
\centerline{\includegraphics[width=1\textwidth]{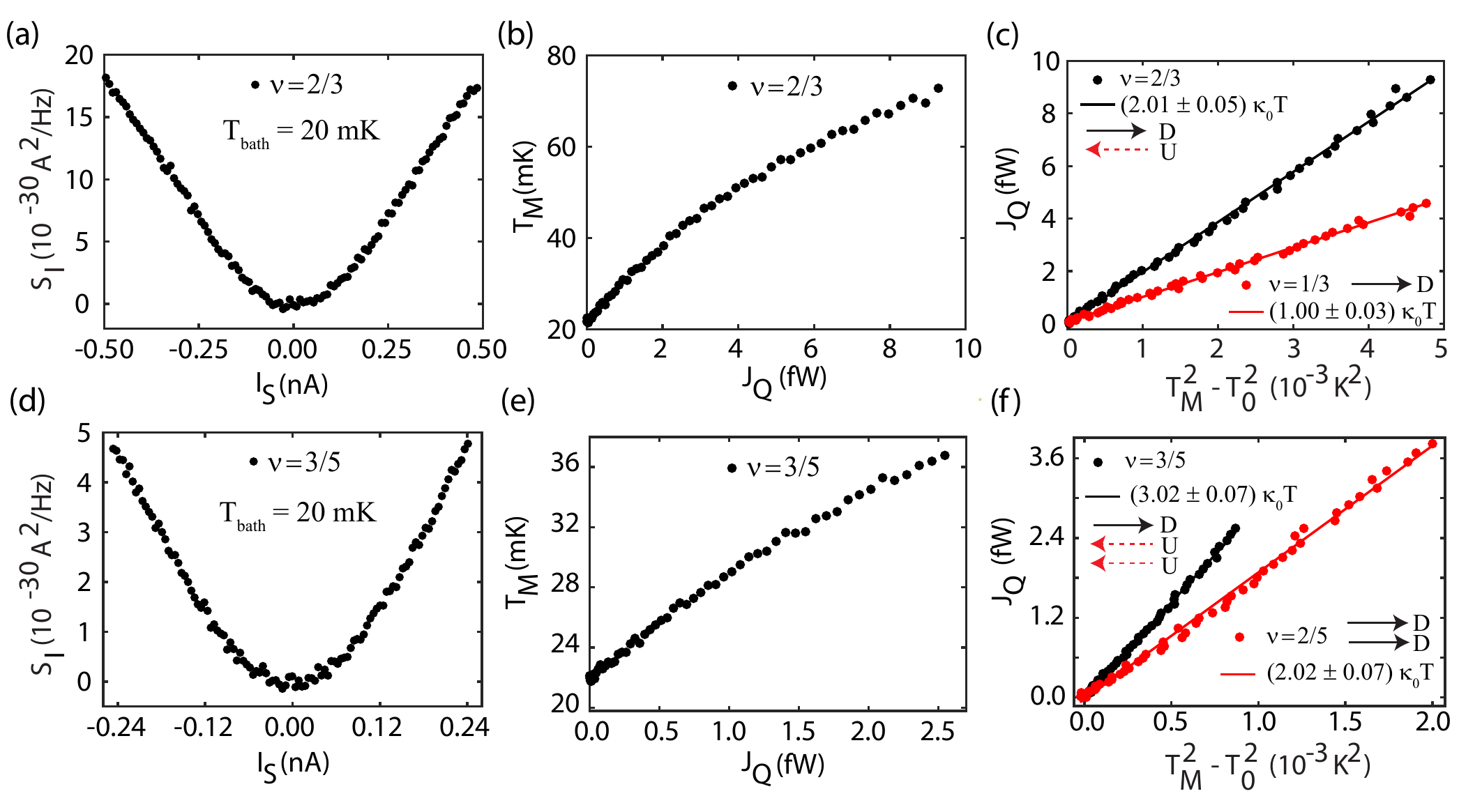}}
\caption{\textbf{Thermal conductance of fractional QH states.} (\textbf{a}) Excess thermal noise $S_{I}$ as a function of source current $I_{S}$ at $\nu=2/3$. The DC currents $I_{S}$ and $-I_{S}$ were injected simultaneously at contacts $S1$ and $S2$, respectively, as shown in Fig. 1d. (\textbf{b}) The temperature $T_{M}$ of the floating contact as a function of the dissipated power $J_{Q}$ at $\nu = 2/3$. (\textbf{c}) $J_Q$ (solid circles) is plotted as a function of $T^2_{M} - T^2_{0}$ at $\nu = $ 2/3 (black) and 1/3 (red). Solid black and red lines are linear fits with $G_{Q} = 2.01\kappa _{0}T$ and $1.00\kappa _{0}T$ for $\nu$ = 2/3 and 1/3, respectively. (\textbf{d}) Excess thermal noise $S_{I}$ as a function of source current $I_{S}$ at $\nu=3/5$. (\textbf{e}) The temperature $T_{M}$ of the floating contact as a function of the dissipated power $J_{Q}$ at $\nu = 3/5$. (\textbf{f}) $J_Q$ (solid circles) is plotted as a function of $T^2_{M} - T^2_{0}$ for $\nu = $ 3/5 (black) and 2/5 (red). Solid black and red lines are linear fits with $G_{Q} = 3.02\kappa _{0}T$ and $2.02\kappa _{0}T$ for $\nu$ = 3/5 and 2/5, respectively . The black and dashed red arrows depict the downstream and upstream modes, respectively, for each edge structure.}
\label{Figure2}
\end{figure*}

\begin{figure*}
\centerline{\includegraphics[width=1\textwidth]{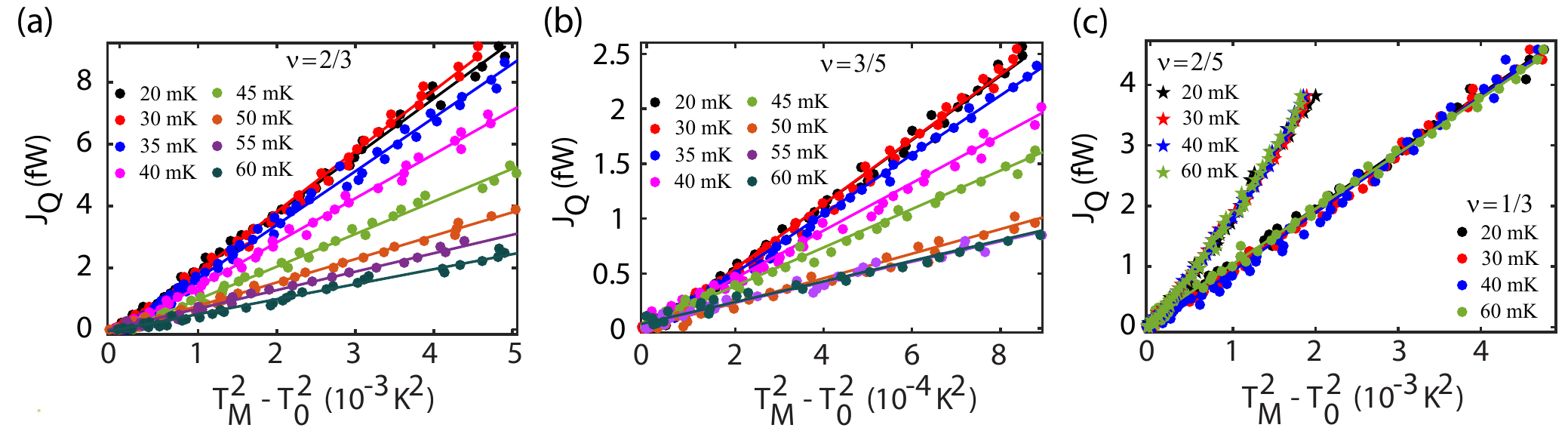}}
\caption{\textbf{Temperature dependence of thermal conductances.} (\textbf{a,b}) $J_Q$ (solid circles) is plotted as a function of $T^2_{M} - T^2_{0}$ at $\nu = $ 2/3 (\textbf{a}) and  $\nu = $ 3/5 (\textbf{b}) at several values of the bath temperature. Solid circles show the experimental data, while solid lines are linear fits to these experimental data points. Different colors  correspond to different bath temperatures as shown in the legend. (\textbf{c}) $J_Q$ (solid circles) is plotted as a function of $T^2_{M} - T^2_{0}$ for $\nu = $ 1/3 ($\bullet$) and  $\nu = $ 2/5 ($\mathbf{\star}$) at several values of the bath temperature. Different colors of the symbols correspond to different bath temperatures, (see legend). For all panels, the thermal conductance $G_Q$ at each temperature is extracted from the slope of the linear fit.} 
\label{Figure3}
\end{figure*}

\noindent\textbf{Thermal conductance measurement:} In contrast to our previous work~\cite{Srivastaveaaw5798,PhysRevLett.126.216803}, to measure the thermal conductance, we simultaneously inject the DC currents $I_{S}$ and $-I_{S}$ at two contacts $S1$ and $S2$, respectively.
Both injected currents flow towards the floating reservoir. This is done in order to keep the potential of the floating contact to be the same as that of all drain contacts.
In this configuration, the dissipated power  at the floating reservoir due to Joule heating is given as $P=\frac{I_S^2}{\nu G_0}$ (see SI). This power dissipation leads to increase of the electron temperature in the floating reservoir. The new steady state temperature $T_{M}$ is determined by the heat balance relation~\cite{jezouin2013quantum,banerjee2017observed,banerjee2018observation,Srivastaveaaw5798,PhysRevLett.126.216803,sivan1986multichannel,butcher1990thermal}
\begin{equation}
P=J_{Q}= J^{e}_{Q}(T_{M},T_{0}) + J^{e-ph}_{Q}(T_{M},T_{0}) = 0.5 N \kappa _{0}(T^{2}_{M}-T^{2}_{0}) + J^{e-ph}_{Q}(T_{M},T_{0}) 
\label{eq.heatbalance}
\end{equation}
Here, $J^{e}_{Q}(T_{M},T_{0})$ is the electronic contribution of the heat current via $N$ chiral edge modes, 
and $J^{e-ph}_{Q}(T_{M},T_{0})$ is the heat loss via electron-phonon cooling. The temperature $T_{M}$ is obtained by measuring the excess thermal noise~\cite{jezouin2013quantum,banerjee2017observed,banerjee2018observation,Srivastaveaaw5798,PhysRevLett.126.216803} along the outgoing edge channels using the Nyquist-Johnson relation
\begin{equation}
S_{I} = \nu k_{B}(T_{M}-T_{0})G_{0} \label{eq.noise}
\end{equation}
For our hBN encapsulated graphite gated devices~\cite{PhysRevLett.126.216803}, the electron-phonon contribution (second term in Eq.~\ref{eq.heatbalance}) was found to be negligible for $T< 100mK$ (see SI). From Eq.~\ref{eq.heatbalance}, one finds $N$, which yields the sought thermal conductance $G_Q=N\kappa_0 T$.

In Fig. 2, we show the detailed procedure to extract the quantized $G_Q$ at the bath temperature $T_0\sim 20 mK$. The measured excess thermal noise $S_{I}$ is plotted as a function of current $I_{S}$ for $\nu =$ 2/3 and 3/5 in Fig.~2(a) and Fig.~2(d), respectively. The resulting heating of the floating reservoir is made manifest by the increase in excess thermal noise with application of the source current $I_S$. The noise and current axes of Fig. 2(a) and 2(d) are converted to $T_{M}$ and $J_{Q}$, yielding Fig.~2(b) for $\nu = 2/3$ and Fig.~2(e) for $\nu = 3/5$, respectively. To extract $G_{Q}$, the heat current $J_{Q}$ is plotted as a function of $T^2_{M} - T^2_{0}$ for $\nu = 1/3$ (red) and 2/3 (black) in Fig.~2(c) and  for $\nu = 2/5$ (red) and 3/5 (black) in Fig.~2(f). The solid circles represent the experimental data, while the solid lines are the linear fits with $G_{Q}=$ $1.00\kappa _{0}T$ (red) and $2.01\kappa _{0}T$ (black) for $\nu= 1/3$ and 2/3, respectively, in Fig.~2(c) and $G_{Q}=$ $2.02\kappa _{0}T$ (red) and $3.02\kappa _{0}T$ (black) for $\nu= 2/5$ and 3/5, respectively, in Fig.~2(f). To further study the temperature dependence of the thermal conductance, $J_{Q}$ is plotted as a function of $T^2_{M} - T^2_{0}$ at several values of the bath temperature for $\nu =$ 2/3 in Fig.~3(a) and for 3/5 in Fig.~3(b). An analogous plot is shown for $\nu=1/3$ (solid circles) and $2/5$ (solid stars) in Fig.~3(c). The slopes of linear fits to the data in these figures allow us to extract the values of $G_Q$. Whereas  the data for the 2/3 and 3/5 states show an explicit dependence of $G_Q$ on bath temperature, the thermal conductance remains independent of the temperature for the 1/3 and 2/5 states, Fig.~3(c).

\noindent\textbf{Results:} In Fig.~4(a), we plot the thermal conductance $G_{Q}$ (extracted from the slope of the linear fits to the data in Fig.~(3) as a function of the bath temperature for $\nu=$ $1/3$ (red), $2/5$ (blue), $2/3$ (magenta), and $3/5$ (black). For completeness, we also show the proposed edge structure of studied FQH states and corresponding theoretically expected $G_{Q}$ values in Fig.~4(b). As evident from the edge structure, for $\nu=$ $1/3$ (red) and $2/5$ (blue), the $G_{Q}$ values should not depend on temperature as there are no CP modes. This is indeed observed in our data shown in Fig.~4(a). On the other hand, for the hole-like $3/5$ state with $N_d = 1$ and $N_u = 2$, we observe the crossover from non-equilibrated regime of $3\kappa_{0}T$ to equilibrated regime of $1\kappa_{0}T$. A similar crossover is observed also for the $2/3$ state with $N_d = N_u = 1$. For this state, at low temperatures, the non-equilibrated regime of $2\kappa_{0}T$ is observed. 
In the equilibrated regime, the transport in this situation ($N_d=N_u$) is diffusive in nature, so that $G_Q$ is expected to tend to zero as $\sim 1/L$ in the long-length limit. Since our device channel length $L$ is limited to $\sim 5\mu m$, we observe a drop of $G_Q$ down to a finite value of $ \sim 0.5 \kappa_{0}T$ at $60mK$. Our result is the first experimental demonstration of the crossover from non-equilibrated to fully equilibrated heat transport in FQH edges with CP modes. Observation of such crossover in a single device has been a long-sought goal, as it can settle the ambiguity in the ground state of the complex even denominator FQH states (e.g., at filling $\nu=5/2$), which remains a subject of active debate.

\begin{figure*}
\centerline{\includegraphics[width=1.0\textwidth]{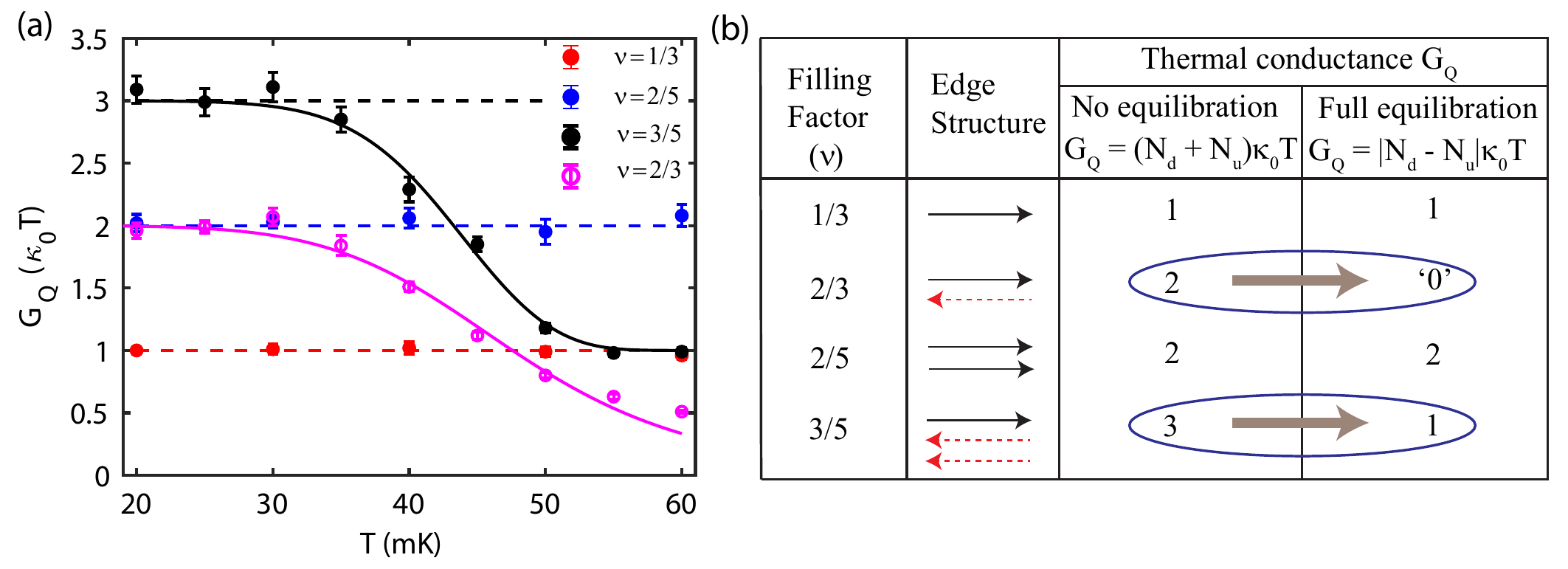}}
\caption{\textbf{Crossover from non-equilibrated to equilibrated heat transport.} (\textbf{a}) Thermal conductance $G_{Q}$, as extracted from the slope of the linear fit in Fig.~3, plotted as a function of the bath temperature for $\nu = $ 1/3 (red), 2/5 (blue), 3/5 (black), and 2/3 (magenta). The horizontal dashed lines correspond to quantized values of $G_{Q}$. The solid curves (black and magenta) are theoretical fits of the data that serve to extract out temperature scaling exponents (see Methods). Error bars correspond to the standard deviation associated with the slope of the linear fit shown in Fig.\ref{Figure3}. (\textbf{b}) Edge structures of the studied FQH states. Solid black and dashed red arrows represent downstream and upstream modes, respectively. The two right-most columns show expected values of the thermal conductance $G_Q$ (in units of $\kappa_0 T$) in the two limiting regimes of the heat transport.}
\label{Figure4}
\end{figure*}

\noindent\textbf{Discussion.} \ 
In this section, we would like to discuss a few additional points related to the expected theoretical regimes of equilibration, the accuracy of our measurement, and temperature exponents of the thermal equilibration lengths. 

\noindent(1) The quantized value $G_Q = (N_d+N_u)\kappa_0 T$ of the thermal conductance in the non-equilibrated regime, $L\ll\ell_{\rm eq}^{H}$, where $\ell_{\rm eq}^{H}$ is the thermal equilibration length, strictly holds if there is no back-scattering of heat at interfaces with contacts. This is fulfilled under an additional condition $L\ll L_T$ where  $L_T\sim T^{-1}$ is the thermal length. In the intermediate regime $L_T\ll L \ll \ell_{\rm eq}^{H}$, a correction to this value is expected to emerge~\cite{Krive1998,protopopov2017,melcer2022}. Thus, the non-equilibrated regime may, in fact, be expected to be split into two plateaus, which is, however, not observed in our experiment.

\noindent(2) The experimental determination of the thermal conductance follows the approach of several preceding works that use two implicit assumptions: (i) current fluctuations propagating from the central contact satisfy the thermal equilibrium distribution, implying the Johnson-Nyquist relation between the contact temperature and the noise; (ii) all power dissipated close to the central contact heats it. When all modes propagate downstream, both these assumptions strictly hold. However, for edges with CP modes, the situation may be somewhat more delicate and some deviations from the assumptions (i) and (ii) may emerge.  This issue was discussed in Ref.~\cite{melcer2022}, where corrections to the procedure of extraction of $G_Q$ were obtained that slightly reduce the experimental value of $G_Q$. We do not include these corrections in the present work. First, they would not affect the identification of the asymptotic regimes. Second, the values of $G_Q$ that we find without including these corrections agree remarkably with the quantized values, both for the non-equilibrated regime (as was also found for bilayer graphene in Ref.~\cite{PhysRevLett.126.216803}) and in the equilibrated limit. It remains to see which features of our device favor this remarkable agreement. We would like to note that the precise determination of $G_Q$ depends on the accuracy of electron temperature and gain of the amplification chain, which are shown in details in SI.

\noindent(3) According to theoretical predictions, the crossover of $G_Q$ between the asymptotic limits of no thermal equilibration ($L \ll \ell_{\rm eq}^H$) and perfect thermal equilibration ($L \gg \ell_{\rm eq}^H$) is described by a function of the dimensionless ratio $L / \ell_{\rm eq}^H$, with the thermal equilibration length scaling as a power of temperature, $\ell_{\rm eq}^H \propto T^{-p}$. Explicit forms of the crossover functions for $\nu=2/3$ and $\nu=3/5$ states are given below in Methods. Our experimental data are well described by these forms. At the same time, the values of the exponent $p$ that are obtained from the fits turn out to be unexpectedly large:  $p=6.3$ for $\nu=2/3$ and $p=9.3$ for $\nu=3/5$, well above $p=2$ expected in the vicinity of the strong-disorder fixed points~\cite{Kane1994,kane1995impurity,protopopov2017}.  This implies that the crossover $G_Q(T)$ is surprisingly sharp as a function of temperature. 
Various mechanisms are known that may in principle lead to large values of $p$ in correlated 1D systems. This may happen if the energy relaxation is controlled by multiparticle processes, or else, by nonlinearities of the edge spectrum. We leave a detailed investigation of this issue in the present context to future research.

\noindent\textbf{Conclusion.} \ 
The findings of this work are a remarkable manifestation of an interplay of equilibration (or absence thereof) and topology in FQH transport.
While the charge transport is in the equilibrated regime, the heat transport crosses over from the non-equilibrated to equilibrated regime, with both asymptotic limits characterized by topologically quantized heat conductances determined by edge quantum numbers. We expect that this physics should be relevant also to other FQH states and materials. 
In particular, interpretation of the experimentally measured thermal conductance  $\frac{5}{2}\kappa_0 T$ at the non-Abelian $\nu=5/2$ state requires assumptions about the presence, absence, or partial character of thermal equilibration~\cite{PhysRevB.99.085309,Ma2020thermal,Simon2020,Asasi2020,park2020noise}. Measurement of the full crossover from the non-equilibrated to equilibrated regime would permit to unambiguously resolve this problem.

\section*{Methods}
\subsection{Device fabrication and measurement scheme:}
In our experiment, encapsulated device (heterostructure of hBN/single layer graphene(SLG)/hBN/graphite) was made using the standard dry transfer pick-up technique~\cite{pizzocchero2016hot}. Fabrication of this heterostructure involved mechanical exfoliation of hBN and graphite crystals on oxidized silicon wafer using the widely used scotch tape technique. First, a hBN of thickness of $\sim 25$ nm was picked up at $90^\circ$C using a Poly-Bisphenol-A-Carbonate (PC) coated Polydimethylsiloxane (PDMS) stamp placed on a glass slide, attached to tip of a home built micromanipulator. This hBN flake was aligned on top of previously exfoliated SLG. SLG was picked up at $90^\circ$C. The next step involved the pick up of bottom hBN ($\sim 25$ nm). This bottom hBN was picked up using the previously picked-up hBN/SLG following the previous process. This hBN/SLG/hBN heterostructure was used to pick-up the graphite flake following the previous step. Finally, this resulting hetrostructure (hBN/SLG/hBN/graphite) was dropped down on top of an oxidized silicon wafer of thickness 285 nm at temperature $180^\circ$C. To remove the residues of PC, this final stack was cleaned in chloroform (\ch{CHCl_3}) overnight followed by cleaning in acetone and iso-propyl alcohol (IPA). After this, Poly-methyl-methacrylate (PMMA) photoresist was coated on this heterostructure to define the contact regions in the Hall probe geometry using electron beam lithography (EBL). Apart from the conventional Hall probe geometry, we defined a region of $\sim$ 5.5 $\mu m^{2}$ area in the middle of SLG flake, which acts as floating metallic reservoir upon edge contact metallization. After EBL, reactive ion etching (mixture of \ch{CHF_3} and \ch{O_2} gas with flow rate of 40 sccm and 4 sccm, respectively at $25^\circ$C with RF power of 60W) was used to define the edge contact. The etching time was optimized such that the bottom hBN did not etch completely to isolate the contacts from bottom graphite flake, which was used as the back gate. Finally, thermal deposition of Cr/Pd/Au (3/12/60 nm) was done in an evaporator chamber having base pressure of $\sim$ $1-2\times 10^{-7}$ mbar. After deposition, a lift-off procedure was performed in hot acetone and IPA. This results in a Hall bar device along with the floating metallic reservoir connected to the both sides of SLG by the edge contacts. The schematics of the device and measurement set-up are shown in Fig. 1(d). The distance from the floating contact to the ground contacts was $\sim$ $5\mu m$ (see SI for optical images). All measurements were done in a cryo-free dilution refrigerator having a base temperature of $\sim 20$mK. The electrical conductance was measured using the standard lock-in technique, whereas the thermal conductance was measured with noise thermometry based on an LCR resonant circuit at resonance frequency $\sim 740$kHz. The signal was amplified by a home-made preamplifier at 4K followed by a room temperature amplifier, and finally measured by a spectrum analyzer. Details of the measurement technique are discussed in our previous work~\cite{Srivastaveaaw5798,PhysRevLett.126.216803} as well as in the SI. 

\subsection{Description of the crossover from the non-equilibrated to equilibrated regime:}

When edge modes are not thermally equilibrated, i.e. for edge lengths $L$ satisfying $L\ll\ell_{\rm eq}^{H}$, the thermal conductance becomes quantized as
\begin{align}
\label{eq:kappa_neq}
    G_Q = (N_d+N_u)\kappa_0 T \,,
\end{align}
which means that every edge mode gives a contribution $1\kappa_0 T$ to $G_Q$. For filling factors $\nu=1/3$, $\nu=2/5$, $\nu=2/3$, and $\nu=3/5$, the corresponding values of the thermal conductance are $G_Q/\kappa_0T = 1$, $2$, $2$, and $3$, respectively.
In fact, the validity of Eq.~\eqref{eq:kappa_neq} requires that $L$ also satisfies $L\ll L_T$, where  $L_T\sim T^{-1}$ is the thermal length. In the intermediate regime $L_T\ll L \ll \ell_{\rm eq}^{H}$, a correction to this value emerges due to back-scattering of heat at interfaces with contacts~\cite{Krive1998,protopopov2017,melcer2022}. For the sake of simplicity, we discard this correction in our analysis in the present work. 

In the regime of full thermal equilibration, $L\gg\ell_{\rm eq}^{H}$, the thermal conductance becomes topologically quantized as
\begin{align}
    \label{eq:kappa_eq}
    G_Q = |N_d-N_u|\kappa_0 T.
\end{align}
For $\nu=1/3$ and $2/5$ we have $N_u=0$, so that Eq.~\eqref{eq:kappa_neq} and~\eqref{eq:kappa_eq} coincide. For such FQH edges, with only downstream modes, the thermal conductance is thus predicted to be $G_Q = N_d \kappa_0 T$, independent of temperature. This is exactly what is observed in our experiment. On the other hand, for FQH edges with CP modes, i.e., with $N_u > 0$, the equilibrated value \eqref{eq:kappa_eq} is smaller than the non-equilibrated value \eqref{eq:kappa_neq}, so that there is a non-trivial crossover of $G_Q$ between the two limits. This is the case for $\nu=2/3$ and $\nu=3/5$. 

 For $\nu=3/5$, we have $N_d = 1$ and $N_u=2$, so that $G_Q/\kappa_0T=1$. It is worth noting that in this case, $N_d-N_u=-1$, implying that the heat flows upstream on the equilibrated edge, i.e., against the charge flow direction. However, the present experimental setup only measures the absolute value of $G_Q$ and does not reveal the heat flow direction on individual edge segments. The crossover function between the non-equilibrated and equilibrated regime is found to be\cite{protopopov2017,Nosiglia2018,PhysRevLett.126.216803}
  \begin{align}
     \frac{G_Q}{\kappa_0 T} = \frac{2+e^{-L/\ell_{\rm eq}^H}}{2-e^{-L/\ell_{\rm eq}^H}} = \frac{2+e^{-kT^p}}{2-e^{-kT^p}},
     \label{eq:crossover-35}
 \end{align}
 where $L/\ell_{\rm eq}^H = kT^p$. Fitting our experimental data to  Eq.~\eqref{eq:crossover-35} with fit parameters $k$ and $p$, we obtain $p \approx 9.34$ (in Fig. 4a).  
 
 For the $\nu=2/3$ state, we have $N_d = N_u =1$, so that the equilibrated limiting value of $G_Q$, Eq.~\eqref{eq:kappa_eq}, is zero. In this case, the crossover takes place between ballistic heat transport in the non-equilibrated regime and heat diffusion in the equilibrated regime\cite{protopopov2017,Nosiglia2018,PhysRevLett.126.216803}:
 \begin{align}
     \frac{G_Q}{\kappa_0 T} =\frac{2\ell_{\rm eq}^H}{L+\ell_{\rm eq}^H} = \frac{2}{1+kT^p}\,.
 \end{align}
 Fitting the experimental data to this form, we get the exponent $p\approx 6.34$ (in Fig. 4a).


\section*{Acknowledgements}
A.D. thanks the Department of Science and Technology (DST) and Science and Engineering Research Board (SERB), India for financial support (DSTO-2051) and acknowledges the Swarnajayanti Fellowship of the DST/SJF/PSA-03/2018-19. S.K.S and R.K. acknowledge Prime Minister's Research Fellowship (PMRF), Ministry of Education (MOE) and Inspire fellowship, DST for financial support, respectively.
 A.D.M. and Y.G. acknowledge support 
by the DFG Grant MI 658/10-2 and by
the German-Israeli Foundation Grant I-1505-303.10/2019. Y.G. acknowledges support by the Helmholtz International Fellow Award, by the DFG Grant RO 2247/11-1, by CRC 183 (project C01), and by the Minerva Foundation. C.S. acknowledges funding from the Excellence Initiative Nano at the Chalmers University of Technology and the 2D TECH VINNOVA competence Center (Ref. 2019-00068). This project has received funding from the European Union's Horizon 2020 research and innovation programme under grant agreement No 101031655, TEAPOT. K.W. and T.T. acknowledge support from the Elemental Strategy Initiative conducted by the MEXT, Japan and the CREST (JPMJCR15F3), JST.

\section*{Author contributions}
S.K.S. and R.K contributed to device fabrication, data acquisition and analysis. A.D. contributed in conceiving the idea and designing the experiment, data interpretation and analysis. K.W and T.T synthesized the hBN single crystals. C.S., A.D.M. and Y.G. contributed in development of theory, data interpretation, and all the authors contributed in writing the manuscript.

\section*{Competing interests}
The authors declare no competing interests.

\section*{Data and materials availability:}
The data presented in the manuscript are available from the corresponding author upon request.

\thispagestyle{empty}
\mbox{}


\section*{Supplementary material}

# Supplementary Information for "Determination of topological edge quantum numbers of fractional quantum Hall phases"

Saurabh Kumar Srivastav[1], Ravi Kumar[1], Christian Spånslätt[2], K. Watanabe[3], T. Taniguchi[3], Alexander D. Mirlin[4,5,6,7], Yuval Gefen[4,8], and Anindya Das[1]*

[1] Department of Physics, Indian Institute of Science, Bangalore, 560012, India.

[2] Department of Microtechnology and Nanoscience (MC2), Chalmers University of Technology, S-412 96 Göteborg, Sweden.

[3] National Institute of Material Science, 1-1 Namiki, Tsukuba 305-0044, Japan.

[4] Institute for Quantum Materials and Technologies, Karlsruhe Institute of Technology, 76021 Karlsruhe, Germany.

[5] Institut für Theorie der Kondensierten Materie, Karlsruhe Institute of Technology, 76128 Karlsruhe, Germany.

[6] Petersburg Nuclear Physics Institute, 188300 St. Petersburg, Russia.

[7] L. D. Landau Institute for Theoretical Physics RAS, 119334 Moscow, Russia.

[8] Department of Condensed Matter Physics, Weizmann Institute of Science, Rehovot 76100, Israel.

---

*anindya@iisc.ac.in

**Section S1: Device fabrication, characterization, and noise measurement setup**

In order to observe well-developed fractional quantum Hall states in graphene, we have used a graphite gated hexagonal boron nitride (hBN) encapsulated graphene (SLG) device (hBN/SLG/hBN/graphite). For the fabrication of the device, we have followed the standard dry transfer pick-up technique [1]. It involves the mechanical exfoliation of hBN and graphite crystals on a oxidized silicon wafer using scotch tape. First, a clean blister free hBN layer of thickness of $\sim$ 25 nm was picked up at $90°C$ using a Poly-Bisphenol-A-Carbonate (PC) coated Polydimethylsiloxane (PDMS) stamp placed on a glass slide, attached to tip of a home build micromanipulator. This hBN flake was aligned on top of previously exfoliated graphene. The graphene was picked up at $90°C$. The next step involved the pick up of bottom hBN ($\sim$ 25 nm). This bottom hBN was picked up using the previously picked-up hBN/SLG following the previous process. The hBN/SLG/hBN heterostructure was used to pick-up a graphite flake following the previous step. Finally, the resulting hetrostructure (hBN/SLG/hBN/graphite) was dropped on top of an oxidized silicon wafer of thickness 285 nm at temperature $180°C$. To remove the residues of PC, the final stack was cleaned in chloroform ($CHCl_3$) overnight followed by cleaning in acetone and and iso-propyl alcohol (IPA). To get the region free from any bubbles and residues, we further performed atomic force microscopy (AFM) topography of the flake. After this, Poly-methyl-methacrylate (PMMA) was coated on the heterostructure to define the contact regions in Hall probe geometry using electron beam lithography (EBL) in clean area of the stack. Apart from the conventional Hall probe geometry, we defined a region of $\sim$ 5.5 $\mu m^2$ area in the middle of the heterostructure, which acts as floating metallic reservoir. After EBL, reactive ion etching (mixture of $CHF_3$ and $O_2$ gas with flow rate of 40 sccm and 4 sccm, respectively at $25°C$ with RF power of 60W) was used to define the edge contacts. The etching time was optimized such that the bottom hBN did not etch completely to isolate the contacts from the bottom graphite flake, which was used as a back gate. Finally, thermal deposition of Cr/Pd/Au (3/12/60 nm) was done in an evaporator chamber having a base pressure of $\sim 1-2 \times 10^{-7}$ mbar. After deposition, a lift-off procedure was performed in hot acetone and IPA. This resulted in a Hall bar device along with the floating metallic reservoir connected to the both sides of SLG by the edge contacts. The AFM topography of stack and the optical image of the full device is shown in Fig. S1(a) and Fig. S1(b), respectively. The distances from the floating contact to the ground contacts was $\sim 5\mu m$. All the measurements are done in a cryo-free dilution refrigerator having a base temperature of $\sim$ 20mK. The electrical conductance was measured using the standard lock-in technique whereas the thermal conductance was measured employing noise thermometry based on an LCR resonant circuit at resonance frequency of $\sim$ 740kHz and amplified by a home made preamplifier at 4K followed by room temperature amplifier, and finally measured by a spectrum analyzer.

Total two-terminal resistances (R) of the device were measured as a function of the bottom graphite

gate voltage ($V_{BG}$) at zero magnetic field. The measured data is fitted with the equation[2–5]

$$R = R_C + \frac{L}{We\mu\sqrt{(n_0^2 + (\frac{C_{BG}(V_{BG}-V_{DP})}{e})^2)}}, \quad (S1)$$

where $R_C$, L, W, $\mu$, and $e$ are, respectively, the contact resistance, length, width, mobility, and electron charge. The carrier concentration of the channel is given by $\frac{C_{BG}(V_{BG}-V_{DP})}{e}$ with $C_{BG}$ and $V_{DP}$ being the capacitance per unit area of the bottom graphite gate, and the voltage at the charge neutrality point, respectively. $n_0$ is the charge inhomogeneity.

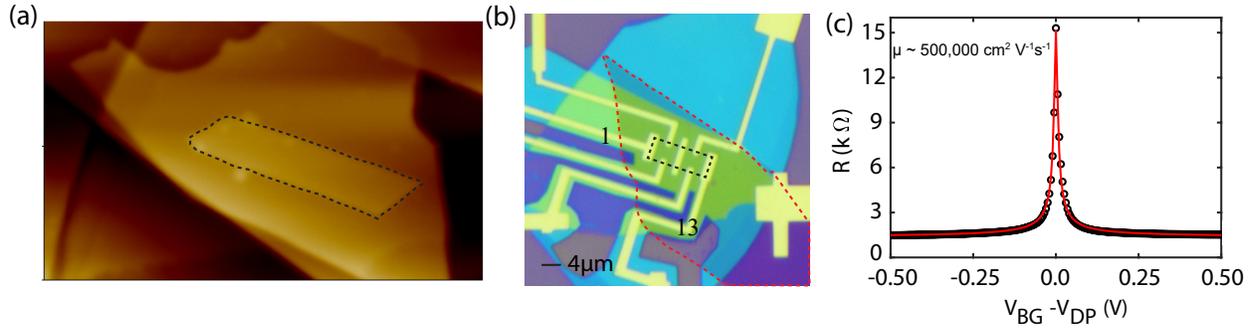

**Fig. S 1: AFM image, optical image, and device characterization.** (**a**) Atomic force microscopic (AFM) topography of the heterostructure. The graphene region is marked with dashed lines. The transport channel was defined in the bubble free region of the graphene flake. (**b**) Optical image of the final device structure. The region of graphene and the bottom graphite are marked by black and red, dashed lines respectively. (**c**) The two-probe gate response measured between contact 1 and 13 (marked in Fig. S1(b)), is plotted as a function of bottom graphite gate voltage at temperature 1.5K. Open circles show the experimental data and the red curve is the fit of data in accordance with Eq. (S1). This fit gives a mobility of $\sim 500,000$ $cm^2V^{-1}s^{-1}$. The high mobility of the device is necessary to observe fractional quantum Hall states. The charge inhomogeneity was found to be on the order of $\sim 2.3 \times 10^9 cm^{-2}$, which is one order of magnitude smaller than $SiO_2$ gated devices.

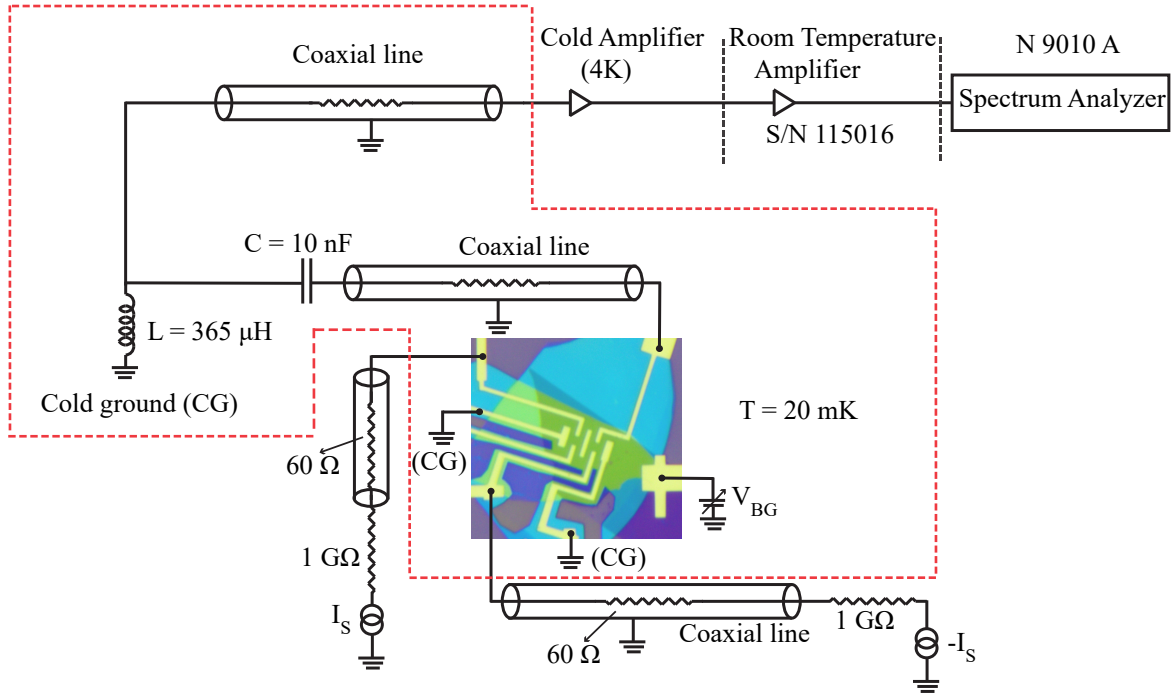

**Fig. S 2: Experimental set-up for noise measurement.** Schematic of the noise measurement set-up. The device was mounted on a chip carrier which was connected to the home made cold finger fixed to the mixing chamber plate of dilution refrigerator. The ground contact (CG) pins are directly shorted to the cold finger to achieve the cold ground. The sample was current biased with currents $+I_S$ and $-I_S$ through two 1 GΩ resistors located at the top of dilution fridge. This was done in order to make the potential of the floating contact to be zero. To measure the temperature of the floating contact, we measure the current fluctuations. Measured current fluctuations are converted on chip into voltage fluctuations using the well defined quantum Hall (QH) resistance $R = h/\nu e^2$, where $\nu$ is the filling factor. This noise signal was amplified with a home made cryogenic voltage pre-amplifier, which was thermalized to 4K plate of dilution refrigerator. This pre-amplified signal was then amplified using a voltage amplifier placed at the top of the fridge at room temperature. After the second stage of amplification, the amplified signal was measured using a spectrum analyzer (N9010A). All the noise measurements were done using the bandwidth $\sim 30$ kHz. The resonant L//C tank circuit was built using an inductor L of $\sim 365$ µH made from a superconducting coil thermally anchored to the mixing chamber plate of the dilution refrigerator. A parallel capacitance C of $\sim 125$ pF develops along the coaxial lines connecting the sample to the cryogenic pre-amplifier. A ceramic capacitance of 10 nF was introduced between sample and inductor to block the DC current along the measurement line.

**Section S2: Gain and Electron temperature calibration:**

The gain of the amplification chain was estimated from temperature dependent Johnson-Nyquist noise (thermal noise)[6]. At zero impinging current, the equilibrium integrated voltage noise spectrum, measured by the spectrum analyser is given as

$$S_V = g^2(4k_B TR + V_n^2 + i_n^2 R^2)BW\,, \qquad (S2)$$

where $g$ is the total gain of amplification chain, $k_B$ the Boltzmann factor, $T$ the temperature, $R$ is the quantum resistance, $V_n^2$ and $i_n^2$ are the intrinsic voltage and current noises of the amplifier, and BW is the frequency bandwidth. At an integer quantum Hall plateau, any change in temperature of mixing chamber (MC) plate will only affect the first term in Eq. (S2), while all other terms are independent of $T$. If one plots $\frac{S_V}{BW}$ as a function of temperature, the slope of the linear curve will be equal to $4g^2 k_B R$. Since at a quantum Hall plateau, the resistance R is exactly known, one can easily calculate the gain of the amplification chain and from the intercept, the intrinsic noises of the amplifier can be found. The gain $g$ can then be calculated using the following equation

$$g = \sqrt{\left(\frac{\partial \left(\frac{S_V}{BW}\right)}{\partial T}\right)\left(\frac{1}{4k_B R}\right)}\,, \qquad (S3)$$

where $\left(\frac{\partial \left(\frac{S_V}{BW}\right)}{\partial T}\right)$ is the slope of the linear fit.

To find the electron temperature at zero impinging current, we measured the integrated voltage noise at resonance frequency for each bath temperature over time and then took the time average of the trace. The averaged integrated voltage noise is given by

$$S_V = g^2(4k_B TR + V_n^2 + i_n^2 R^2)BW\,. \qquad (S4)$$

Since we have already estimated the gain (from the slope) and the intrinsic noise of amplification chain (from the intercept) (see caption of Fig. S3), the corresponding electron temperature $T_0$ at base temperature of the mixing chamber plate can be found directly from the known value of the measured noise at zero bias. $T_0$ is given by

$$T_0 = \frac{\left(\left(\frac{S_V}{g^2 BW}\right) - (V_n^2 + i_n^2 R^2)\right)}{4k_B R}\,. \qquad (S5)$$

The estimated electron temperature at several values of the bath temperature is shown in Fig. S3(e) in table form.

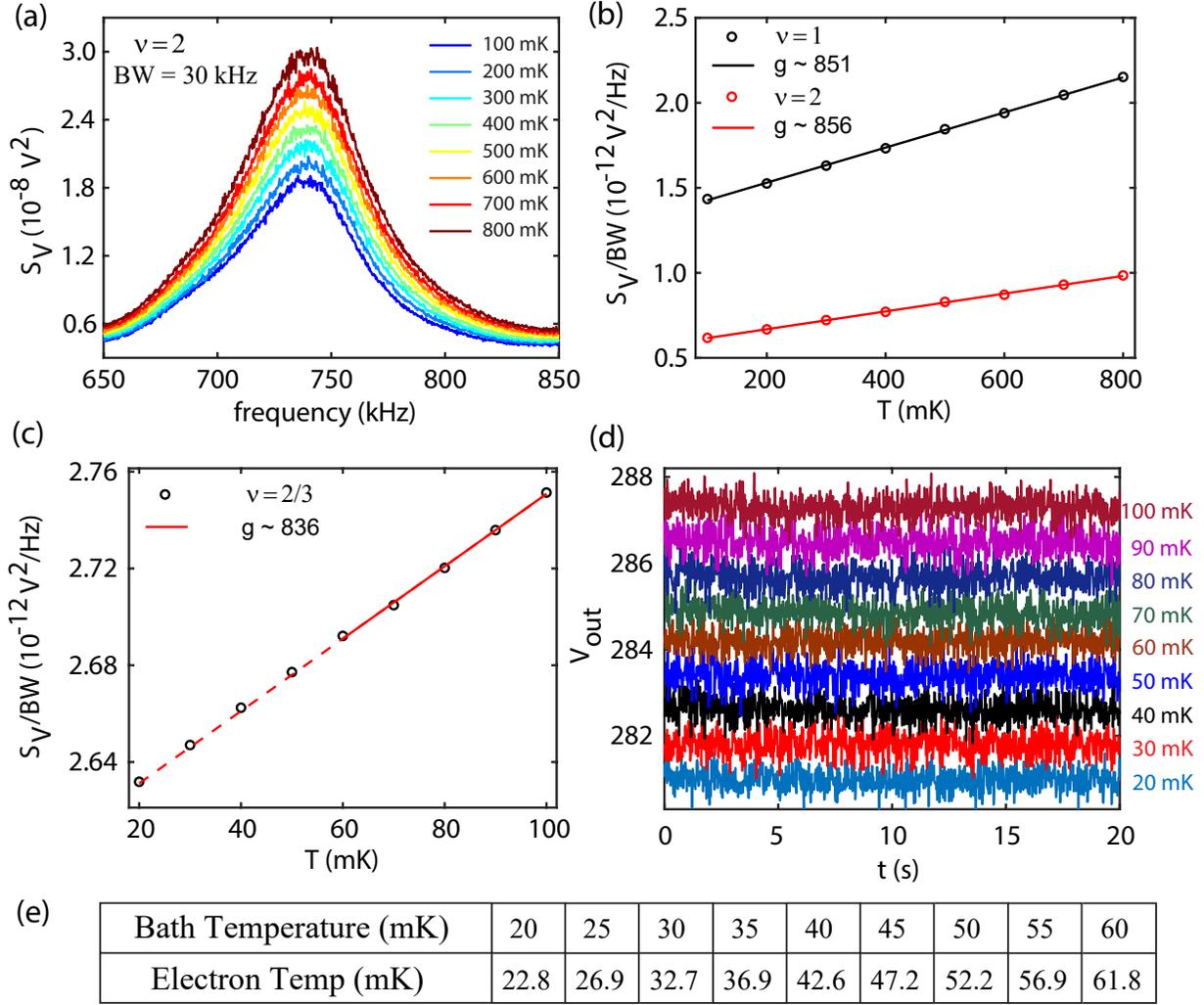

| Bath Temperature (mK) | 20 | 25 | 30 | 35 | 40 | 45 | 50 | 55 | 60 |
|---|---|---|---|---|---|---|---|---|---|
| Electron Temp (mK) | 22.8 | 26.9 | 32.7 | 36.9 | 42.6 | 47.2 | 52.2 | 56.9 | 61.8 |

**Fig. S 3: Gain of the amplification chain and electron temperature at zero impinging current.** (**a**) The integrated voltage noise measured at zero bias is plotted as a function of frequency at different temperatures at $\nu = 2$. From these plots, the resonance frequency of the tank circuit was found to be $\sim$740 kHZ. (**b**) Open circles represent integrated noises divided by bandwidth ($S_V/BW$) at resonance frequency as a function of temperature for $\nu = 1$ (black) and $\nu = 2$ (red), respectively. Solid lines are the linear fit of these data points. Using Eq. (S3) and the slope information from these linear fits, the calculated gain was found to equal $\sim$851 ($\nu = 1$) and $\sim$856 ($\nu = 2$). (**c**) Symbols represent the plot of $S_V/BW$ as a function of temperature at $\nu = 2/3$ at low temperatures. The solid red line is the linear fit of the data and from the slope of this line, the calculated gain was found to equal $\sim$836. (**d**) Each trace represents the measured output voltage $V_{out}$ after the second stage of the amplification by the spectrum analyser at $\nu = 2/3$ for several values of the bath temperature shown by different colours. Each trace curve is the average of 200 scans. This measured output voltage is related to the voltage noise $S_V$ via the relation $S_V = V_{out}^2$. The electron temperature was calculated using Eq. (S5) (**e**) Table of bath temperature and corresponding estimated electron temperatures.

6

**Section S3: Joule heating and temperature ($T_M$) of the floating reservoir:**

We have used two different configurations to achieve a hot metallic floating contact. In configuration 1, the metallic island remains at finite potential while in configuration 2, its potential is identically zero: the potential of the ground contacts. The noise data presented in the main manuscript is obtained using configuration 2. Here, we find the equations relating the dissipated power and the injected currents in both configurations.

**Finite potential of floating contact.** The current injection schematic is shown in Fig. S4(a). In this configuration, the floating reservoir reaches a new equilibrium potential $V_M = \frac{I_S}{2\nu G_0}$ with the filling factor $\nu$ of graphene determined by $V_{BG}$. The potential of the $S$ contact is $V_S = \frac{I_S}{\nu G_0}$. The power input to the floating reservoir is then $P_{in} = \frac{1}{2}(I_S V_S) = \frac{I_S^2}{2\nu G_0}$, where the pre-factor $\frac{1}{2}$ results due to the fact that equal power dissipates at the source and the floating reservoirs. Similarly, the outgoing power from the floating reservoir is $P_{out} = \frac{1}{2}(2 \times \frac{I_S}{2} V_M) = \frac{I_S^2}{4\nu G_0}$. Thus, the resulting injected power dissipation in the floating reservoir due to joule heating is $P_{in} - P_{out} = \frac{I_S^2}{4\nu G_0}$.

An alternative way to quantify the dissipation in floating contact is to calculate the power dissipation at the hot spots[7]. Whenever there is a change in the potential near contacts, hot spots will generate heat. There will be two hot spot located near the floating contact, two near cold ground contacts and one at the back of the source contacts as show in Fig. S4(a). The power dissipated at the floating contact equals the sum of the power dissipated at the two hot spots formed near the floating contact. The half of the injected power from the source contact will drop at the back of the source contact and other half will be equally drop at four other hot spots, out of which two are formed near the floating contact and two near the cold ground contacts. The power dissipated near the floating contact is then $2 \times \left(\frac{1}{4}\left(\frac{I_S^2}{2\nu G_0}\right)\right) = \frac{I_S^2}{4\nu G_0}$.

**Zero potential of floating contact.** In this configuration, currents $+I_S$ and $-I_s$ are injected from two contacts as shown schematically in Fig. S4(b). This leads to a zero potential of the floating contact. In this configuration, two hot spots form near the floating contact and two other are formed at the back of the source contacts as shown in Fig. S4(b). Then, the dissipated power will be $2 \times \left(\frac{I_S^2}{2\nu G_0}\right) = \frac{I_S^2}{\nu G_0}$.

**Electron temperature of the floating contact.** The resulting increase in the electron temperature ($T_M - T_0$) of the floating contact is determined from the generated excess thermal noise [8–12]: $S_I = 2G^* k_B (T_M - T_0)$ with $\frac{1}{G^*} = \frac{1}{G_L} + \frac{1}{G_R}$, where $G_L$ and $G_R$ are the conductance of left and right channel respectively. In our device structure, we have $\frac{1}{G^*} = \frac{1}{\nu G_0} + \frac{1}{\nu G_0}$, hence $S_I = \nu k_B (T_M - T_0) G_0$.

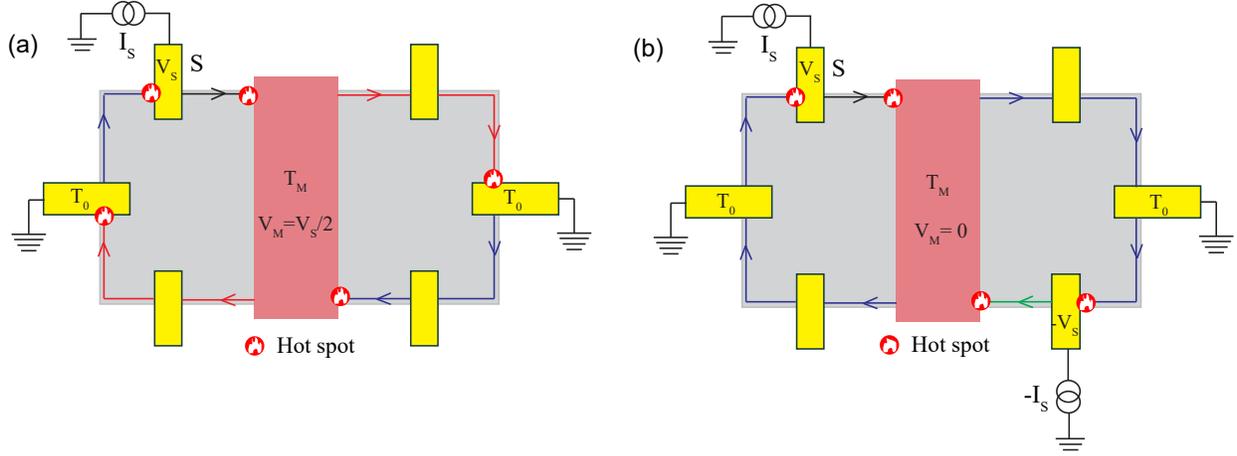

**Fig. S 4: Current source configurations and corresponding hot spot positions.** **(a)** Schematic of the current injection and hot spot positions in configuration 1. Here, the floating contact remains at finite potential. The current $I_S$ injected from source contact $S$ is carried along the quantum Hall edge channel to the floating contact. From the floating contact, the current splits into two equal parts, which propagate towards two cold grounds. In this scenario, the potential of the floating contact becomes half of the potential of the source contact. The power dissipation in this configuration is $J_Q = \frac{I_S^2}{4\nu G_0}$. **(b)** Configuration 2, which leads to zero potential of the floating contact. Current $+I_S$ and $-I_S$ are injected simultaneously from two diagonal contacts. In this configuration, no hot spots form near the cold ground contacts. The power dissipation in this configuration is $\frac{I_S^2}{\nu G_0}$.

**Section S4: Equipartitioning of the current and robustness of fractional plateaus.**

The equipartitioning of the injected current is crucial for the thermal conductance measurement. In other words, the bulk filling fraction on both side of the floating contact should be the same. This should be verified in order to rule out any possibility of bulk contributions as well as to ensure the validity of the dissipated power relation to the source current $I_S$. To verify this, we use a measurement configuration shown in Fig. S5(a). Current is injected from the contact source S and the voltage is measured at contact S, R, and T. At the quantum Hall plateaus, the measured voltage at R and T contact is found to be half of the voltage measured at contact S, which establishes the equipartitioning of the current. In Fig. S5(b), resistances measured at different contacts are plotted as functions of the back gate voltage. Here, the resistance is obtained by dividing the measured potential at different contacts by the injected current. Since the measured voltage at contact R and T is half of the voltage measured at contact S, the resistance value shown in Fig. S5(b) at these contacts are found to be half of the original quantum resistance.

In addition to the equipartitioning, it is important that the fractional states remain robust at the max-

imum electron temperature ($\sim 100$) mK reached in our measurement. To check that, we measure the resistance in a configuration as shown in Fig. S5(c), which encodes the longitudinal resistance. In this configuration, the current $i_R$ is injected from contact R. The clockwise chirality ensures that the injected current terminates at the cold grounds at any QH plateaus. The resistance in this configuration, plotted in Fig. S5(d), has the same properties as a longitudinal resistance: in the absence of bulk transport, the voltage $V_S$ is determined by the equilibrium potential of the ground contact. The observation of the vanishing resistance plateaus at 20 mK (black) and 100 mK (red) further supports the robustness of the FQH states.

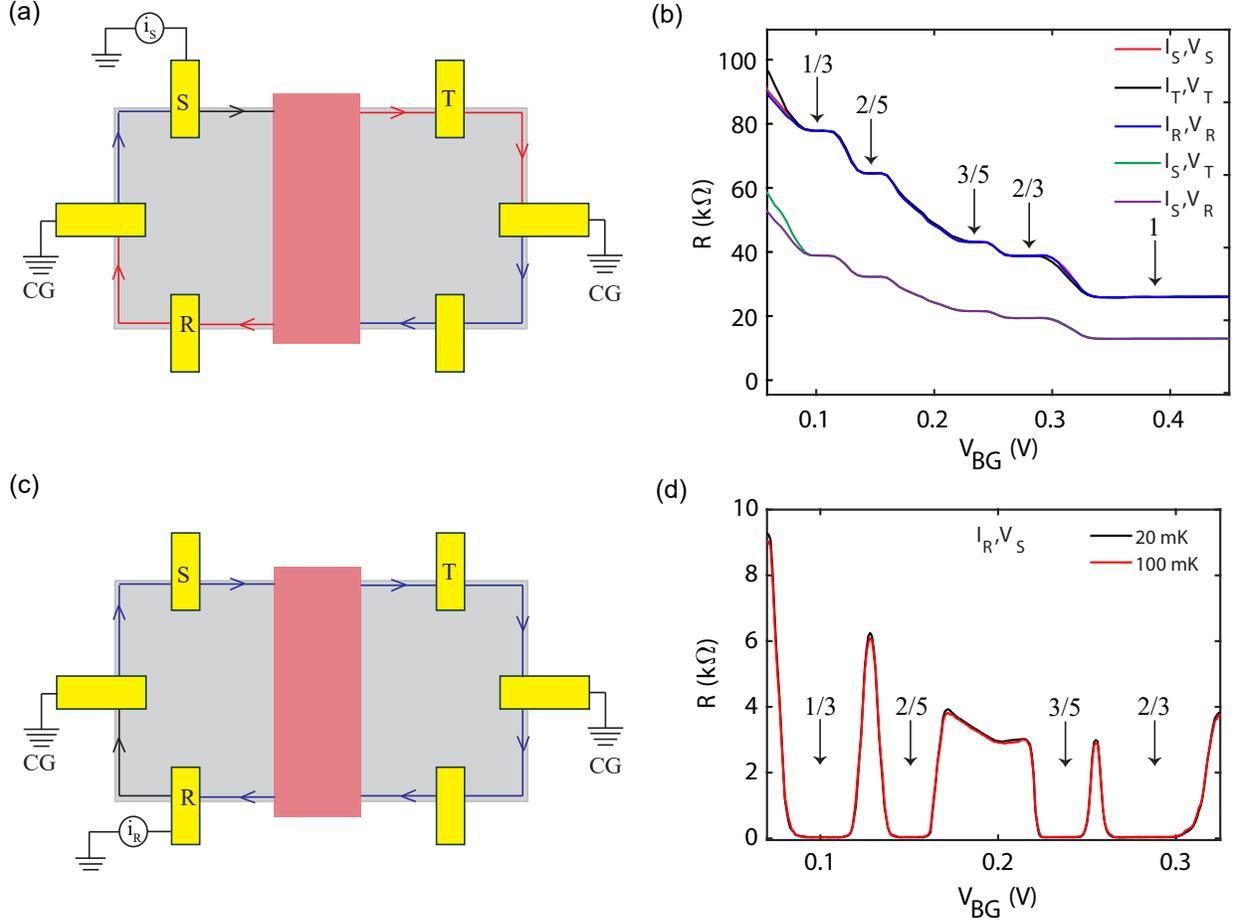

**Fig. S 5: Equipartitioning of the current.** (a) Schematic of the measurement configuration. Current $i_S$ is injected from source contact $S$, and the voltage is measured at contacts $S$, $R$, and $T$. (b) The resistance $R$ ($V/i_S$) measured at different contacts are plotted as functions of the gate voltage. Different colours represents traces taken at different contacts. In the legend, the subscript of the $I$ and $V$ shows the current injection and voltage probe contact respectively. At quantum Hall plateaus, the resistance curves taken at the contacts $R$ (purple) and $T$ (green) (injecting current at $S$) are lying on top of each other, which demonstrates that the filling on both side of the floating constant are equal. The halving of the magnitude of the resistance measured at these contacts establish that the injected current is equally split into two parts from the floating contact. (c) Schematic of the longitudinal resistance measurement configuration. Current $i_R$ is injected from contact $R$, and the voltage is measured at contact $S$. The resistance measured in this configuration encodes the longitudinal resistance. (d) The resistance $R$ ($V/i_R$) measured at contact $S$ is plotted as function of the gate voltages. Black and red colours represent traces taken for 20 mK and 100 mK of the bath temperature, respectively.

10

## Section S5: Values of the electrical conductance without charge equilibration.

To estimate the electrical conductance values of hole-conjugate fractional quantum Hall states in the absence of charge equilibration along the propagation length for our device configuration, we follow a Landauer-Büttiker approach [13]. In this calculation, we assume full charge equilibration at the Ohmic contacts including the floating metallic contacts. We calculate the electrical conductance at $\nu = 2/3$ and $\nu = 3/5$ edges, which are believed to host counter propagating bare charge modes in absent of the equilibration. The schematic of the device with contact number is shown in Fig. S6.

**The $\nu = 2/3$ edge:** The multiprobe device geometry is shown in Fig. S6(a). In the absence of charge equilibration, the edge structure of the $\nu = 2/3$ state consists of a downstream charge mode of charge $e$ and an upstream mode of charge $-e/3$. For a multi-probe device, the net current flowing in $i^{th}$ contact is given by

$$I_i = \sum_j (G_{j \leftarrow i} V_i - G_{i \leftarrow j} V_j), \tag{S6}$$

where $G_j \leftarrow i$ is the conductance from the $i^{th}$ contact to the $j^{th}$ contact and $V_i$ is the voltage of the $i^{th}$ contact. In matrix form, we then have

$$\begin{pmatrix} I_1 \\ I_2 \\ I_3 \\ I_4 \\ I_5 \\ I_6 \\ I_7 \end{pmatrix} = \frac{e^2}{h} \begin{pmatrix} 1+1/3 & -1/3 & 0 & 0 & 0 & 0 & -1 \\ -1 & 2(1+1/3) & -1/3 & 0 & -1 & -1/3 & 0 \\ 0 & -1 & 1+1/3 & -1/3 & 0 & 0 & 0 \\ 0 & 0 & -1 & 1+1/3 & -1/3 & 0 & 0 \\ 0 & -1/3 & 0 & -1 & 1+1/3 & 0 & 0 \\ 0 & -1 & 0 & 0 & 0 & 1+1/3 & -1/3 \\ -1/3 & -0 & 0 & 0 & 0 & -1 & 1+1/3 \end{pmatrix} \begin{pmatrix} V_1 \\ V_2 \\ V_3 \\ V_4 \\ V_5 \\ V_6 \\ V_7 \end{pmatrix}. \tag{S7}$$

Contacts 4 and 7 are taken as grounded, and we may eliminate the corresponding entries from the matrix (S7). After eliminating the rows and column associated with contacts 4 and 7, we get the reduced equation

$$\begin{pmatrix} I_1 \\ I_2 \\ I_3 \\ I_5 \\ I_6 \end{pmatrix} = \frac{e^2}{h} \begin{pmatrix} 1+1/3 & -1/3 & 0 & 0 & 0 \\ -1 & 2(1+1/3) & -1/3 & -1 & -1/3 \\ 0 & -1 & 1+1/3 & 0 & 0 \\ 0 & -1/3 & 0 & 1+1/3 & 0 \\ 0 & -1 & 0 & 0 & 1+1/3 \end{pmatrix} \begin{pmatrix} V_1 \\ V_2 \\ V_3 \\ V_5 \\ V_6 \end{pmatrix}. \tag{S8}$$

Since current is injected only at contact 1, the current column reduces to

$$\begin{pmatrix} I_1 \\ I_2 \\ I_3 \\ I_5 \\ I_6 \end{pmatrix} = \begin{pmatrix} I \\ 0 \\ 0 \\ 0 \\ 0 \end{pmatrix}, \tag{S9}$$

11

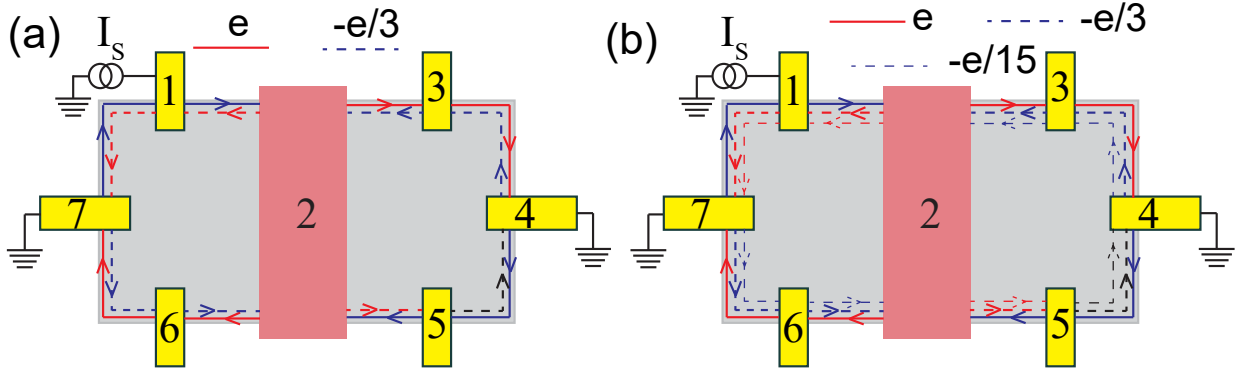

**Fig. S 6: Schematic for the electrical conductance calculation using Landauer Büttiker Formalism.** The contacts are marked by the numbers. The charged mode with charge $e$ is shown by solid line and charge mode with charge $e/3$ is shown by dashed line.

and we can solve for the voltages. We then find

$$\begin{pmatrix} V_1 \\ V_2 \\ V_3 \\ V_5 \\ V_6 \end{pmatrix} = I \frac{h}{e^2} \begin{pmatrix} 0.8625 \\ 0.4500 \\ 0.3375 \\ 0.1125 \\ 0.3375 \end{pmatrix}. \tag{S10}$$

Voltage measured at contact 1 is therefore

$$V_1 = 0.8625 \times I \frac{h}{e^2} \tag{S11}$$

and the conductance at source contact 1 is

$$G = \frac{I}{V_1} = 1.16 \frac{e^2}{h}. \tag{S12}$$

Similarly, the conductances at contacts 3 and 6 become

$$G = \frac{I}{V_3} = 2.96 \frac{e^2}{h} \tag{S13}$$

and, respectively,

$$G = \frac{I}{V_6} = 2.96 \frac{e^2}{h}. \tag{S14}$$

**The $\nu = 3/5$ edge:** The multiprobe device geometry is shown in Fig. S6(b). In the absence of charge equilibration, the edge structure of the $\nu = 3/5$ state consists of one downstream charge mode of charge $e$

12

and two upstream mode with charges $-e/3$ and $-e/15$. Following the same approach as for the $\nu = 2/3$ state, we here get the equation system (assuming contact 4 and 7 are grounded)

$$\begin{pmatrix} I_1 \\ I_2 \\ I_3 \\ I_5 \\ I_6 \end{pmatrix} = \frac{e^2}{h} \begin{pmatrix} 1+\frac{1}{3}+\frac{1}{15} & -\frac{1}{3}-\frac{1}{15} & 0 & 0 & 0 \\ -1 & 2(1+\frac{1}{3}+\frac{1}{15}) & -\frac{1}{3}-\frac{1}{15} & -1 & -\frac{1}{3}-\frac{1}{15} \\ 0 & -1 & 1+\frac{1}{3}+\frac{1}{15} & 0 & 0 \\ 0 & -\frac{1}{3}-\frac{1}{15} & 0 & 1+\frac{1}{3}+\frac{1}{15} & 0 \\ 0 & -1 & 0 & 0 & 1+\frac{1}{3}+\frac{1}{15} \end{pmatrix} \begin{pmatrix} V_1 \\ V_2 \\ V_3 \\ V_5 \\ V_6 \end{pmatrix}.$$
(S15)

Current is only injected at contact 1, i.e.,

$$\begin{pmatrix} I_1 \\ I_2 \\ I_3 \\ I_5 \\ I_6 \end{pmatrix} = \begin{pmatrix} I \\ 0 \\ 0 \\ 0 \\ 0 \end{pmatrix},$$
(S16)

and the voltages become

$$\begin{pmatrix} V_1 \\ V_2 \\ V_3 \\ V_5 \\ V_6 \end{pmatrix} = I\frac{h}{e^2} \begin{pmatrix} 0.8374 \\ 0.4310 \\ 0.3079 \\ 0.1232 \\ 0.3079 \end{pmatrix}.$$
(S17)

The voltage measured at contact 1 reads

$$V_1 = 0.8375 \times I\frac{h}{e^2},$$
(S18)

so that conductance at source contact 1 equals

$$G = \frac{I}{V_1} = 1.19\frac{e^2}{h}.$$
(S19)

Similarly, the conductances at contacts 3 and 6 become

$$G = \frac{I}{V_3} = 3.25\frac{e^2}{h}$$
(S20)

and, respectively,

$$G = \frac{I}{V_6} = 3.25\frac{e^2}{h}.$$
(S21)

The conductances are summarized in Table S1. The calculated conductances for the $\nu = 2/3$ edge was found to be $2.16\frac{e^2}{h}$ and $2.96\frac{e^2}{h}$ at source(S) and reflected/transmitted (R/T) contacts, respectively. Similarly, for the $3/5$ edge, the calculated conductances were found to be $1.19\frac{e^2}{h}$ and $3.25\frac{e^2}{h}$ at source(S) and

13

reflected/transmitted (R/T) contacts, respectively. However, in our experiment, the measured values of the conductance at source ($I_S/V_S$) and reflected/transmitted ($I_S/V_R$ or $I_S/V_T$) contacts were found to be $0.67\frac{e^2}{h}$ and $1.33\frac{e^2}{h}$ for $\nu = 2/3$, and $0.60\frac{e^2}{h}$ and $1.20\frac{e^2}{h}$ for $\nu = 3/5$, respectively. These measured values suggest that the charge equilibration of the counter propagating edge modes along the propagation length is well established in our device.

| Filling Factor ($\nu$) | Calculated conductance in absence of charge equilibration ($I/V_i$) (in $e^2/h$) | | | Experimentaly measured conductance ($I/V_i$) (in $e^2/h$) | | |
| --- | --- | --- | --- | --- | --- | --- |
| | Source (1) | Reflected (2) | Transmitted (3) | Source (1) | Reflected (2) | Transmitted (3) |
| 2/3 | 1.16 | 2.96 | 2.96 | 0.67 | 1.33 | 1.33 |
| 3/5 | 1.19 | 3.25 | 3.25 | 0.60 | 1.20 | 1.20 |

**Table S1: Comparison of calculated electrical conductances using the Landauer- Büttiker formalism and our measured conductances.** The electrical conductances calculated for the hole-like states assuming no charge equilibration between counter propagating edge channels is always much larger than the experimentally measured values, which are instead agreement with values expected for fully equilibrated edges.

**Section S6: Extraction of averaged noise data from raw data.**

The excess thermal noise data presented in the main manuscript in Fig. 2(a) and 2(d) and other excess thermal noise plots presented in this supplementary file is the averaged data of several experimental traces of raw data. This is explicitly demonstrated in Fig. S(7) for $\nu = 2$. The blue scan shown in Fig. S7(a) is a single trace measured by the spectrum analyser. As is clearly seen, the data points are highly fluctuating and it is extremely difficult to do any quantitative analysis from this data. The red curve shown in Fig. S7(b) is the average of ∼1000 such raw data traces. We follow the same procedure for all fractional fillings considered in this work. After getting this averaged data for the excess thermal noise, we extract $J_Q$ and $T_M$ from this data using the equations obtained in section S3. In Fig. S7(b), the green curve shows the extracted data from Fig. S7(a), while the black solid circles are the 9 point average of the corresponding green curve.

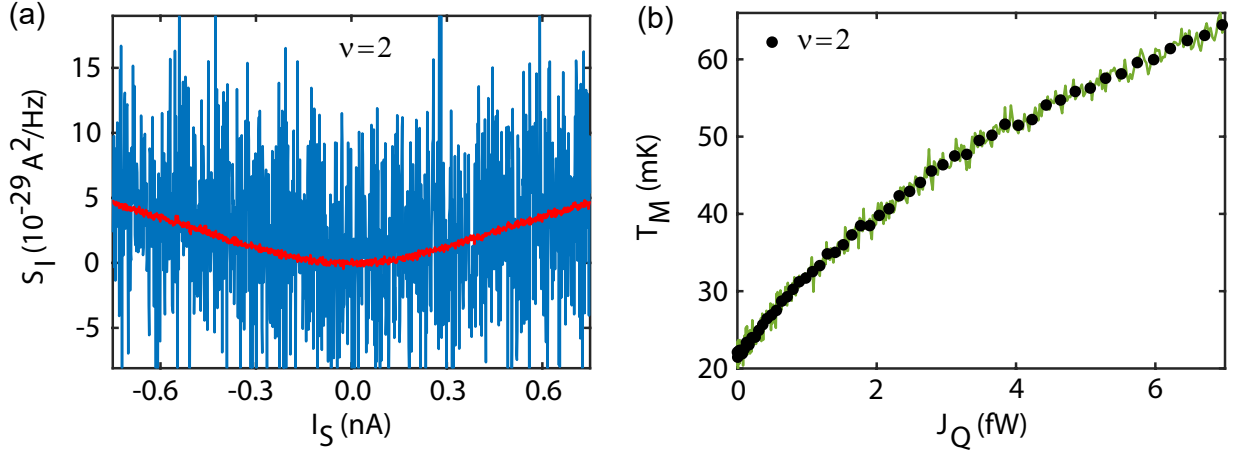

**Fig. S 7: Extraction of average data from raw data** (a) Excess noise for a single scan (blue) and the average of 1000 scans (red) at $\nu = 2$. (b) The solid, green curve shows the data extracted directly from raw excess thermal noise data at $\nu = 2$. Solid, black circles display 9 point averages of the corresponding raw data.

**Section S7: Extraction of the thermal conductance for the finite potential configuration of the floating contact.**

In the main manuscript, all thermal noise data points shown are taken at zero potential of the floating contact. It is done by injecting currents $+I_S$ and $-I_S$ simultaneously from two contacts as shown schematically in Fig. 1(b) (main manuscript) and in Fig. S4(b). We repeat the thermal conductance measurement using our conventional measurement configuration used in our previous work. In this measurement configuration, the current is only injected from one contact as shown schematically in Fig. S4(a), which leads to the finite potential of the floating contact at finite bias. The measured value of the thermal conductance in this configuration matches with the zero potential configuration of the floating contact shown in main manuscript.

**Thermal conductance measurement of integer quantum Hall state:** In this section, we discuss the measurement of the thermal conductance for integer fillings $\nu = 1$, 2 and 3, in the finite potential configuration of the floating contact. This is shown schematically in Fig. S8(a). The current is injected from the contact $S$ and the resulting excess thermal noise is measured at contact T. The excess thermal noise is plotted as function of the current $I_S$ in Fig. S8(b),8(c), and 8(d) for $\nu = 1$, 2, and 3, respectively. The $S_I$ and $I_S$ axis from these plots are converted into $T_M$ and $J_Q$, respectively, which is further plotted in Fig. S8(e) for $\nu = 1$ (black), 2 (red), and 3 (blue). To extract the values of $G_Q$, $J_Q$ is plotted as function of $T_M^2 - T_0^2$ for $\nu = 1$ (black), 2(red), and 3(blue). The solid circles represent the experimental data, while the solid curves are the linear fits of this data, resulting in $G_Q = 1.03$, 1.99, and 2.99 $\kappa_0 T$ for $\nu = 1$, 2, and 3, respectively.

15

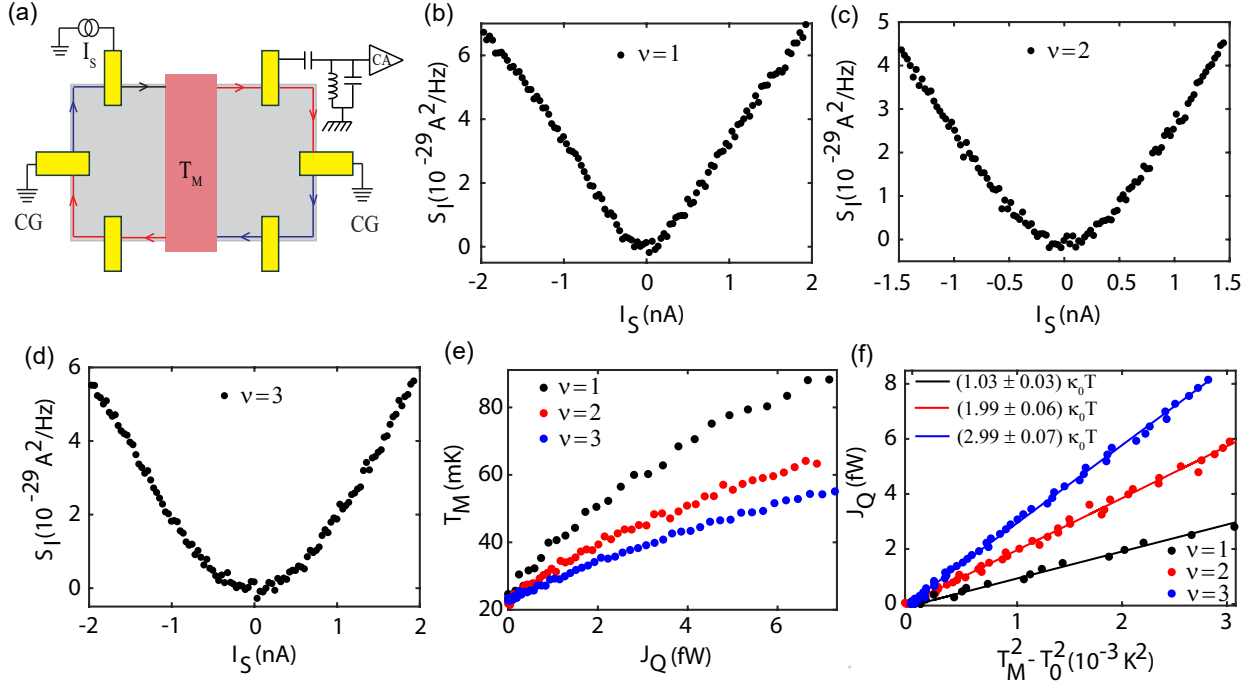

**Fig. S 8: Thermal conductance measurement of integer quantum Hall states** (a) Schematic of the measurement configuration. (b) Excess thermal noise $S_I$ is plotted as a function of source current $I_S$ at filling $\nu = 1$ (b), 2 (c), and 3 (d). (e) Temperature $T_M$ (extracted from the excess thermal noise shown in (b,c,d)) of floating contact is plotted as a function of dissipated power $P = J_Q$ (obtained using $P = \frac{I_S^2}{4\nu G_0}$ and the heat balance equation) for filling factors $\nu = 1$ (black), 2 (red), and 3 (blue), respectively. Symbols display the extracted temperature data using equation $S_I = \nu k_B (T_M - T_0) G_0$. (f) $J_Q$ plotted as a function of $T_M^2 - T_0^2$. Solid circles display the data and the solid lines are the linear fits of these data points with $G_Q = 1.03$, 1.99, and 2.99 $\kappa_0 T$ for $\nu = 1$, 2, and 3, respectively.

**Thermal conductance measurement of fractional quantum Hall states** Similar to the integer quantum Hall states, the current is injected from the contact $S$ and the resulting excess thermal noise is measured at contact T. The excess thermal noise is plotted as function of the current $I_S$ in Fig. S9(b), 9(c), 9(d), and 9(e) for $\nu = 1/3$, 2/5, 3/5, and 2/3, respectively. The $S_I$ and $I_S$ axis from these plots are converted into $T_M$ and $J_Q$, respectively, which is further plotted for $\nu = 1/3$ (red) and 2/3 (black) in Fig. S9(f), and for $\nu = 2/5$ (red) and 3/5 (black) in Fig. S9(g). To extract the value of $G_Q$, $J_Q$ is plotted as functions of $T_M^2 - T_0^2$ for $\nu = 1/3$ (red) and 2/3 (black) in Fig. S9(h), and for $\nu = 2/5$ (red) and 3/5 (black) in Fig. S9(i). The solid circles represent the experimental data while the solid curves are the linear fits of these data points, resulting in $G_Q = 1.02$, 2.04, 2.02, and 3.05 $\kappa_0 T$ for $\nu = 1/3$, 2/3, 2/5, and 3/5, respectively. All these data points were taken at bath temperature 30 mK.

16

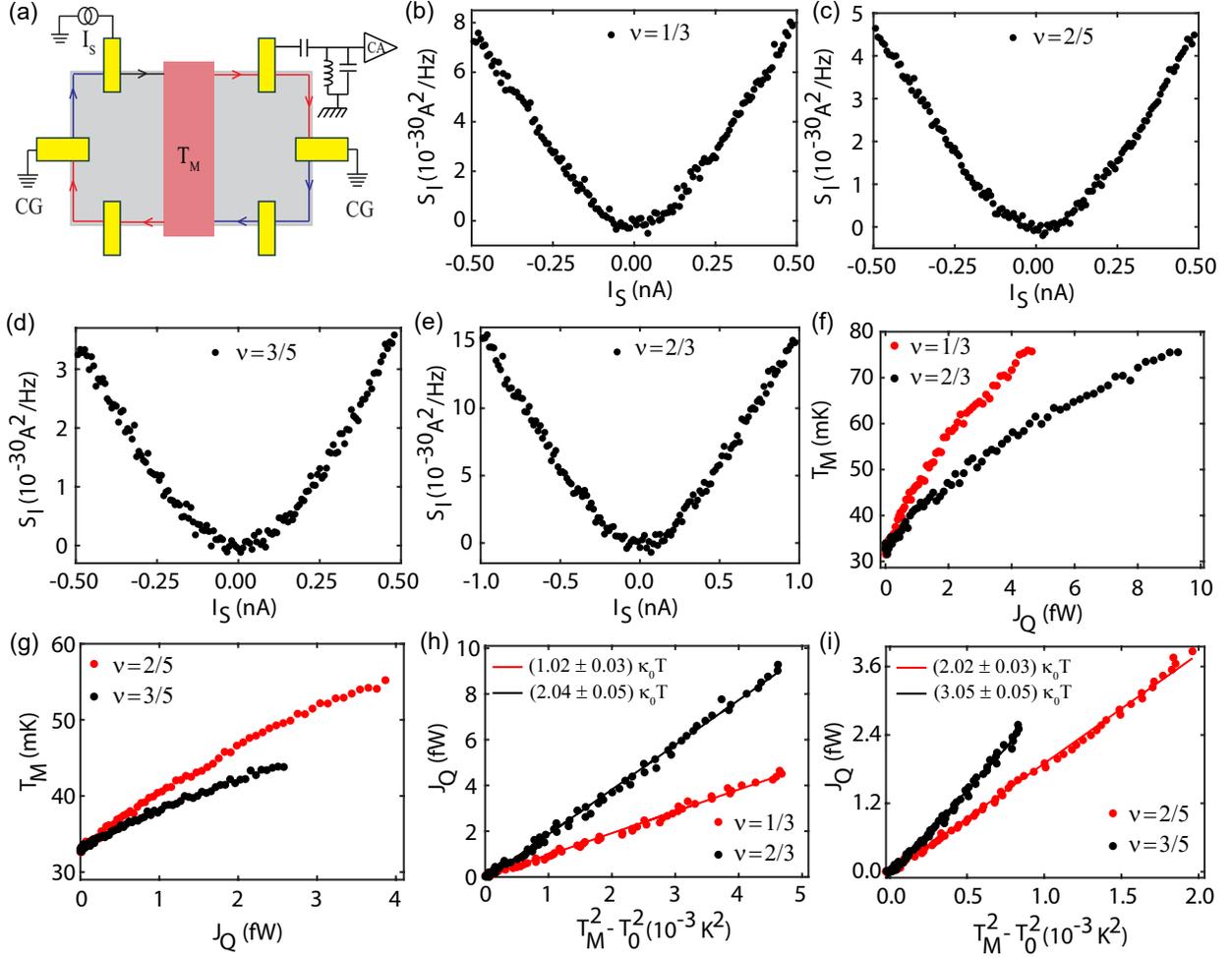

**Fig. S 9: Thermal conductance measurement for fractional quantum Hall states** (**a**) Schematic of the measurement configuration. (**b**) Excess thermal noise $S_I$ is plotted as a function of the source current $I_S$ at fillings $\nu = 1/3$. (**c-e**) same as for panel (**b**) but for fillings 2/5, 3/5, 2/3. (**f, g**) Floating island temperature $T_M$ (extracted from the excess thermal noise shown in (b,c,d,e)), plotted as a function of the dissipated power $P = J_Q$ (obtained using $P = \frac{I_S^2}{4\nu G_0}$ and the heat balance equation) for $\nu = 1/3$ (red) and 2/3 (black) in Fig. S9(f), and for $\nu = 2/5$ (red) and 3/5 (black) in Fig. S9(g). Symbols display the extracted temperature data using equation $S_I = \nu k_B (T_M - T_0) G_0$. (**h, i**) $J_Q$ plotted as a function of $T_M^2 - T_0^2$ for $\nu = 1/3$ (red) and 2/3 (black) in Fig. S9(h), and for $\nu = 2/5$ (red) and 3/5 (black) in Fig. S9(i). Solid circles display the data and the solid lines are the linear fits of these data points with $G_Q = 1.02, 2.04, 2.02,$ and $3.05$ $\kappa_0 T$ for $\nu = 1/3, 2/3, 2/5,$ and 3/5, respectively. All these data were taken at bath temperature of 30 mK. The $G_Q$ of particle like (1/3, 2/5) states matches with expected value of $N_d \kappa_0 T$, while for hole like (2/3, 3/5) states, it matches with the non-equilibrated values of $(N_d + N_d) \kappa_0 T$. Here, $N_d$ and $N_u$ are the number of downstream and upstream modes, respectively.

17

## Section S8: Heat loss by electron-phonon Cooling.

The heat balance equation (1) in the main manuscript, contains, in addition to the electronic contribution, a mechanism of heat transfer via electron-phonon cooling ($J_Q^{e-ph}$). To estimate the contribution of $J_Q^{e-ph}$, we have subtracted the electronic contribution ($J_Q^e$) from the total heat current ($J_Q$),

$$J_Q^{e-ph} = J_Q - J_Q^e. \tag{S22}$$

We have $J_Q^e = 0.5 N \kappa_0 (T_M^2 - T_0^2)$, where $N$ is the total number of electronic channels leaving the floating contact. Usually, $J_Q^{e-ph}$ has the functional form of $J_Q^{e-ph} = \beta(T_M^q - T_0^q)$. In our devices, $J_Q^{e-ph}$ was found to be negligible below $\sim 100 mK$. It can be seen from Fig. S-10a that the deviation from linearity happens beyond $\sim 100 mK$. The power exponent ($q$) was found to be 5 in this case as shown in Fig. S-10b. It should be mentioned that $q$ was found to be varying in the range between 4 and 6 in our earlier work[14] and elsewhere[15]. Although the deviation from linearity shown in Fig. S-10a is consistent with the heat loss due to electron-phonon cooling, other possible mechanisms of heat loss can not be ruled out completely.

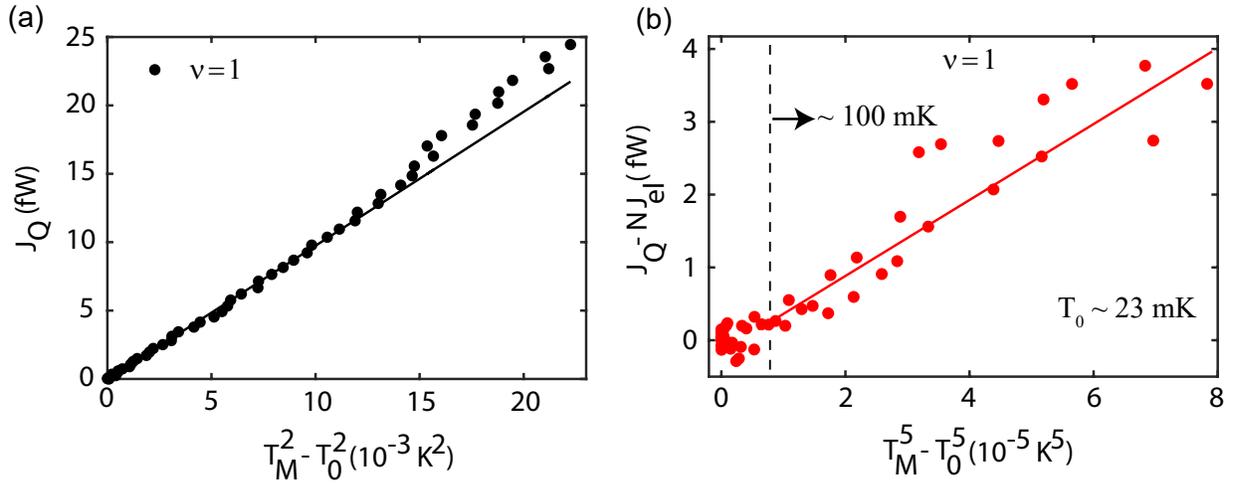

**Fig. S 10: Heat loss by electron-phonon cooling.** (a) $J_Q$ plotted as function of $(T_M^2 - T_0^2)$ for $\nu = 1$. The solid circles represents the experimental data while the solid curve shows the theoretical line corresponding to the electronic contribution of heat flow. It can be seen that the experimental data points start deviating from the theoretical line beyond $\sim 15$ fW. This deviation is attributed to heat losses by electron-phonon cooling. (b) Solid circles display the electron-phonon contribution of the heat loss ($J_Q^{e-ph}$) as a function of $(T_M^5 - T_0^5)$ for $\nu = 1$. The solid line is a linear fit with slope $\beta \sim 0.053 nW/K^5$. The contribution of $J_Q^{e-ph}$ is only seen beyond $\sim 100$ mK, as indicated by the vertical, dashed line

**Section S9: Contact resistance and source noise.**

The finite contact resistance of the source contact can create additional unwanted noise. To estimate this noise, we first need to have some estimate of the contact resistance. In our case, the contact resistance was extracted from the low-frequency resistance data. As shown in Fig. S5(a), a current $i_S$ was injected at source contact and the voltage was measured at the same contact. This voltage probe measures the voltage $i_S.(R_0 + R_L + R_C)$ with $R_0$ the quantum Hall resistance, $R_C$ the contact resistance, and $R_L$ the line resistance. The measured resistance will then be $R_0 + R_L + R_C$. We have measured the line resistance $R_L$ to 265 Ω separately. After subtracting the line resistance, the conductance will be equal to $\frac{1}{R_0+R_C}$. Hence, the transmittance $t$ will be given by $t = \frac{R_0}{R_0+R_C}$. Once the transmittance is known, one can determine the source noise $2eI(1-t)$. Since the amplifier is situated in the path of right moving edge channels in one arm of the device, it will measure only part of the generated source noise. In particular, for our device configuration, the amplifier will always measure only a $\frac{1}{4}$th of the source noise. The estimated contact resistance, transmittance and the source noise is shown in Table S2. The reflection coefficient was always less than 0.25% for all fractional filling factors. The source noise measured by the amplifier would therefore be at least 2-3 orders of magnitude smaller than the measured excess thermal noise.

| Filling Factor($v$) | Measured resistance ($R_0+R_L+R_C$) (Ω) | Line resistance $R_L$ (Ω) | Contact Resistance ($R_0+R_L+R_C$) - $R_L$ - $R_0$ in (Ω) | Trans-mittance ($t$) | Source Noise / 4 ($10^{-30}$A$^2$/Hz) at |
|---|---|---|---|---|---|
| 1/3 | 77888 | 265 | 184 | 0.9976 | 0.016 @ $I_{max}$ = 0.25 nA |
| 2/5 | 64866 | 265 | 68 | 0.9989 | 0.004 @ $I_{max}$ = 0.25 nA |
| 3/5 | 43353 | 265 | 66 | 0.9985 | 0.006 @ $I_{max}$ = 0.25 nA |
| 2/3 | 39029 | 265 | 45 | 0.9988 | 0.008 @ $I_{max}$ = 0.25 nA |

**Table S2: Contact resistance and source noise.** Measured and estimated values for the contact resistance, the transmittance, and the maximum source noise for fractional filling factors.



\end{document}